\documentclass[12pt]{iopart}
\usepackage{graphicx}
\usepackage{dcolumn}
\usepackage{bm}
\usepackage{amssymb}
\usepackage[T1]{fontenc}
\usepackage[latin9,utf8]{inputenc}
\usepackage{amssymb}
\usepackage[colorlinks]{hyperref} 
\usepackage{braket}
\usepackage{placeins}
\usepackage{cleveref}
\usepackage{xcolor}
\begin{document}
	
\title[Mesoscopic Interference for Metric and Curvature (MIMAC)]{Mesoscopic Interference for Metric and Curvature (MIMAC) \& Gravitational Wave Detection}

\author{Ryan J. Marshman$^1$, Anupam Mazumdar$^2$,
Gavin W. Morley$^3$, Peter F. Barker$^1$, Steven Hoekstra$^2$ and Sougato Bose$^1$}
\address{$^1$ Department of Physics and Astronomy, University College London, Gower Street, WC1E~6BT London, United Kingdom.}
\address{$^2$ Van Swinderen Institute, University of Groningen, 9747 AG Groningen, The Netherlands.}
\address{$^3$ Department of Physics, University of Warwick, Gibbet Hill Road, Coventry CV4~7AL, United Kingdom.}

\begin{abstract}
	A compact detector for space-time metric and curvature is highly desirable. Here we show that quantum spatial superpositions of mesoscopic objects could be exploited to create such a detector. We propose a specific form for such a detector and analyse how asymmetries in its design allow it to directly couple to the curvature. Moreover, we also find that its non-symmetric construction and the large mass of the interfered objects, enable the detection gravitational waves (GWs). Finally, we discuss how the construction of such a detector is in principle possible with a combination of state of the art techniques while taking into account the known sources of decoherence and noise. To this end, we use Stern-Gerlach (SG) interferometry with masses $\sim 10^{-17}$kg, where the interferometric signal is extracted by measuring spins and show that accelerations as low as $5\times10^{-15}\textrm{ms}^{-2}\textrm{Hz}^{-1/2}$, as well as the frame dragging effects caused by the Earth, could be sensed. The GW sensitivity scales differently from the stray acceleration sensitivity, a unique feature of the proposed interferometer. We identify mitigation mechanisms for the known sources of noise, namely Gravity Gradient Noise (GGN), uncertainty principle and electro-magnetic forces and show that it could potentially lead to a meter sized, orientable and vibrational noise (thermal/seismic) resilient detector of mid (ground based) and low (space based) frequency GWs from massive binaries (the predicted regimes are similar to those targeted by atom interferometers and LISA).
\end{abstract}

\submitto{\NJP}

\section{Introduction \label{sec:Introduction}}
Matter wave interferometry has been very successful with atoms \cite{RevModPhys.81.1051}, and implemented already with macromolecules ($10^4$ amu mass) \cite{gerlich2007kapitza}. There has been a push to extend this to larger superpositions, or more macroscopic masses~\cite{bose1999scheme,armour2002entanglement,marshall2003towards,sekatski2014macroscopic,romero2010toward,romero2011large,khalili2010preparing,scala2013matter,bateman2014near,yin2013large,pino2018chip,clarke2018growingPublished,ringbauer2016generation,khosla2018displacemon,kaltenbaek2012MAQRO}, or both~\cite{wan2016free,romero2017coherent} to explore collapse model modifications of quantum mechanics~\cite{penrose1996gravity, bassi2013models} and to test whether the gravitational field is fundamentally quantum in nature ~\cite{Bose:2017nin,Marletto2017}. However, as it will be a considerable effort to realize these interferometers, it is really important to examine their usefulness beyond the purely fundamental and postulated processes. In addition, while searching for applications, it makes sense to be optimistic about the regimes achievable by combining several state-of-the-art quantum technologies and experimental techniques. With the above motivations, here we examine sensor/detector applications of the large mass, large superposition regime~\cite{wan2016free,romero2017coherent,Bose:2017nin} in interferometry. We find an application in which such superpositions are used to detect fully the classical gravitational effects in a location as quantified by the metric and curvature. This comes against a backdrop of proposals of smaller particle interferometers \cite{Colella1975,Anandan1984,Mcguirk2002} or larger quantum optomechanical systems \cite{qvarfort2018gravimetryPublished,Armata2017} to detect a $g_{00}$ metric component, whose variations can be used to infer the associated component of curvature, the direct measurement of such curvature~\cite{asenbaum2017phase} or to detect the Earth's rotation~\cite{WSC(1979),Anandan1977} or general relativistic effects~\cite{Dimopoulos2008a,GravityProbeB,roura2017circumventing,roura2018gravitational}. The most challenging entities to detect are the gravitational waves (GWs), the $g_{ij}$ metric components, whose detection has been a huge recent success using kilometre long optical interferometers ~\cite{Abbott2016, Harry2010}, with future devices proposed in space~\cite{LISA}. On the other hand there are also proposals for usage of atomic interferometers~\cite{Chiao2004,Roura2006,Foffa2006,Dimopoulos2008,Dimopoulos2009,MIGA,MAGIS} and various resonators \cite{Raetzel2017,Raetzel2018,Geraci2013,pontin2018levitated,ando2010torsion}, but nothing yet on the potential of interferometers for propagating (untrapped) objects much larger than single atoms.

In this paper, we will discuss how mesoscopic-object interference could be employed for detecting metric and curvature (MIMAC), and moreover present an example scheme based on the Stern-Gerlach principle \cite{wan2016free,Folman2013,Folman2018,folman2019}. Thus this paper has three aims: 1. To show that large mass interferometry with a certain asymmetric design would allow the capability to directly detect unprecedented regimes of inertial and gravitational effects with compact sized devices, 2. To present an explicit example interferometer and demonstrate how to, in general, infer the signals it can sense 3. Present a viability study for this specific interferometer to highlight that these regimes will soon be accessible. Here it noteworthy that there could be other, perhaps more viable schemes based on other methods to prepare superpositions of mesoscopic objects. Such investigations would be fuelled by our findings under aims 1 and 2. 

 In the particular type of interferometer which we study as an example model, although a spatial interferometry involving superpositions of separated motional states takes place, the output signal of the interferometer is encoded in a spin degree of freedom in a manner which is insensitive to the initial noise in the motional state (thermal and seismic).  We demonstrate that it can be used to observe the metric and, as a result of using a non-symmetric set-up, also ``directly'' observe the derivatives in the interferometric signal which determine the curvature of a perturbed Minkowski metric  (as opposed to indirectly inferring the curvature by measuring the metric in nearby locations and then approximating derivatives of the metric). It is due to this ability to directly sense curvature through the interferometer that we describe the interferometer as sensitive to metric {\emph{and}} curvature (CF Section~\ref{sec:Observability of Space-time}). Additionally, these interferometers enable the measuring of the Earth's frame dragging and gravitational waves of certain strength and frequency range. In all these cases, it is remarkable, and indeed directly due to the high masses of the objects undergoing interferometry, that the interferometer is very compact (one meter or smaller), and highly sensitive at a single object level, i.e., does not require a high flux of objects.
	
This paper will proceed as follows: Section~\ref{sec:Non-relativistic action} will review the general form of the action for a mass moving through non-trivial space-time in the non-relativistic limit. It also presents the standard arguments in favour of using larger ``mesoscopic" masses as the interferometric particles. Of course the observations of this section are independent of the specific type of mesoscopic object interferometer that one uses, and as such is adaptable to other future proposals. Section~\ref{sec:Interferometric setup} presents a specific proposal for a mesoscopic object interferometer for detecting the space-time metric and its curvature (MIMAC). This interferometer employs Stern-Gerlach interferometry and is a modified version of the previously proposed interferometer suggested in other contexts with both atoms~\cite{Folman2013,Folman2018,folman2019} and mesoscopic particles~\cite{wan2016free,Bose:2017nin}. Section~\ref{sec:Observability of Space-time} will present the exact components of the space-time metric detectable by the suggested form of MIMAC in such a way to also provide a guide to analyse future interferometer proposals. Sections~\ref{sec:Newtonian potential},~\ref{sec:Frame dragging} and~\ref{sec:Gravitational Waves} will present and discuss how the most interesting signals found in Section~\ref{sec:Observability of Space-time}, namely Newtonian gravity and its associated curvature, frame dragging and gravitational waves (GW), can be detected. This will include suggesting the basic experimental parameters required for detection and presenting the resulting sensitivities. Finally Section~\ref{sec:Practical Impelementation} will discuss in detail the requirements for the most challenging of the signals, gravitational wave observation in the mid-band frequency and demonstrate how, although ambitious, such a device does not appear to be beyond realisability. This is done by presenting how current state-of-the-art techniques match or beat the minimum experimental requirements for theoretical gravitational wave observation. We also discuss the primary expected noise sources and their effects in such a device, namely decoherence effects, gravity gradient noise, the Heisenberg uncertainty limit and electro-magnetic effects. While Section~\ref{sec:Non-relativistic action} points out the potential of a new regime and Section~\ref{sec:Interferometric setup} is presents a necessary modification of an existing apparatus, Sections~\ref{sec:Newtonian potential}-\ref{sec:Gravitational Waves} are entirely new theoretical results. Section~\ref{sec:Practical Impelementation} is, of course, compiling state-of-art commercially available equipment and experimental achievements by various laboratories to justify the potential realisability of our scheme.

\section{Non-relativistic action \label{sec:Non-relativistic action}}

The signal extracted by an interferometer coupled to the space-time metric is the phase difference ($\Delta \phi$) between the two arms of the interferometer. This is given by $\Delta \phi = \Delta S/\hbar$, where $\Delta S$ is the difference in action between the two paths through the interferometer. As such, relative to any classical gravimeter or similar classical experiment, this $1/\hbar$ dependence in the final phase will hugely amplify the final measured signal in a quantum interferometer. If we consider the space-time metric, $g_{\mu\nu}$, as slightly perturbed, as is true for earth based measurements, the space-time metric can be written as $g_{\mu\nu}=\eta_{\mu\nu}+h_{\mu\nu}$ where $\eta_{\mu\nu}$ is the standard Minkowski metric with signature $\left(-+++\right)$ and $h_{\mu\nu}$ is some small perturbation that may have space and time dependencies. We will also take the non-relativistic limit for the interferometric particles motion, as a result the laboratory time $t$ can be taken to be approximately equivalent to the proper time. Then the action for a particle of mass $m$ travelling along a trajectory $\iota$ in the is

\begin{eqnarray}
S&=&-mc\int_{\iota}\rmd s \nonumber\\
&=&-mc\int_{\iota} \sqrt{\rmd s^2} \nonumber\\
&=&-mc\int_{\iota} \left(g_{\mu\nu}\frac{\rmd x^{\mu}}{\rmd\tau}\frac{\rmd x^{\nu}}{\rmd\tau}\right)^{1/2}\rmd\tau\nonumber\\
&=& mc^2\int_{\iota} \biggl[-\left(\eta_{00}+h_{00}\right)\frac{\rmd t^2}{\rmd\tau^2} - h_{0j}\frac{v^{j}}{c}\frac{\rmd t}{\rmd\tau} - h_{i0}\frac{v^{i}}{c}\frac{\rmd t}{\rmd\tau} - \left(\delta_{ij} + h_{ij}\right)\frac{v^{i}}{c}\frac{v^{j}}{c}\biggr]^{1/2}~\rmd\tau\qquad \\	 
&\approx& m\int_{\iota} \left[c^2\left(1-\frac{h_{00}} {2}\right) - ch_{0j}v^{j} - \left(\delta_{ij}+h_{ij}\right)\frac{v^{i}v^{j}}{2}\right]~\rmd t\, \label{eq:General action formula}
\end{eqnarray} 
where $\delta_{ij}$ is the Kronecker delta. From the above formula it is evident that compared to the $h_{00}$ component (Newtonian potential), the terms  $h_{0j}$ (Frame Dragging) are harder to detect as $c$ is replaced by a non-relativistic velocity $v_j$, while $h_{ij}$ (Gravitational Waves) will be the most difficult to detect with $c^2$ replaced by $v_i v_j$.  On the other hand, a high value of $m$ (compared to atomic masses) are expected to increase the sensitivity to all terms, potentially allowing the detection of signals which would otherwise be to small to see. For example, for similar velocities and times (we will show how to achieve this in later sections), when one uses nano-objects of mass $10^{-17}$ kgs, there is an O($10^{8}$) times amplification in the final signal ($\Delta \phi$) compared to a heavy atom.

\section{Interferometric Setup \label{sec:Interferometric setup}}

\begin{figure}
	\centering
	\includegraphics[width=0.8\linewidth]{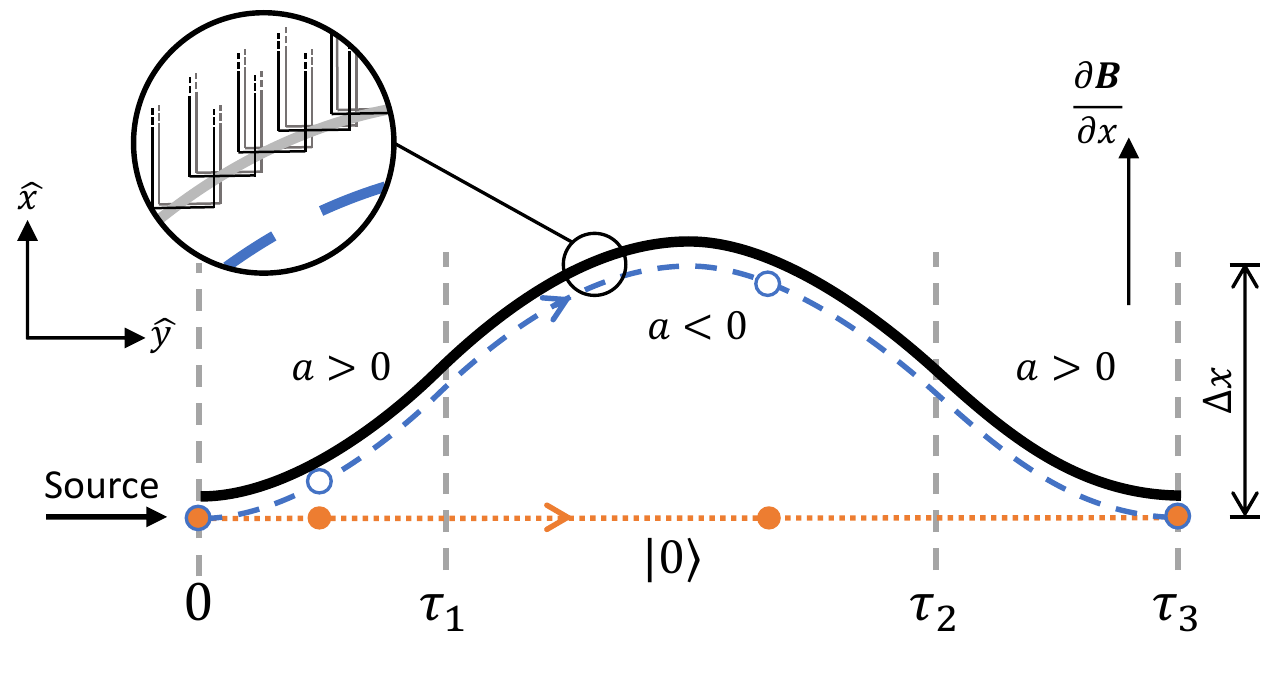}
	\caption{Interferometer path diagram showing spin $\left|\pm1\right\rangle$ dashed blue path and spin $\left|0\right\rangle$ path dotted in orange. The magnetic field source (thick black line) could be shaped to follow the non-spin-zero path such that it can provide a large magnetic field gradient without a needing an exceedingly large magnetic field. The detail cut-away shows how this can be sourced by many flat current carrying wires arranged to approximate the ideal curved shape for the magnetic field source. By running a current through each pair of wires in the same direction only when the particle is directly below it  we can ensure the particle experiences an approximately uniform magnetic field gradient only in the desired direction. The maximum superposition of $\Delta x=a\tau_1^2$ is achieved halfway through the interferometry process. The vertical dotted lines show the position when the acceleration direction changes occur. The circular cut-out shows the detail for how the magnetic field source could actually be implemented as many individual flat current carrying wires turned on in sequence.  Note the unusual axis orientation with the $x$ axis vertical representing the spatial superposition distance \label{fig:Interferometer path diagram}}
\end{figure}

Here we will present an example form for a mesoscopic object interferometer. This proposal amounts to a modification of the devices previously proposed for atomic and mesoscopic interferometry. Stern-Gerlach interferometry of the type we are proposing to use requires a spin embedded in a nano-crystal. This is a very generic requirement and the proposal does not rely on a specific type of spin system or crystalline host. The primary requirement is that a superposition of embedded spin states remains coherent for the duration of an experiment, which is a common requirement in the field of quantum computing with spin qubits. For the moment, consider a mesoscopic mass (nano-meter sized crystal) containing an embedded spin $1$ degree of freedom (three spin states $|+1\rangle,|0\rangle,|-1\rangle$). One example is a diamond crystal of nanometer scale diameter with a nitrogen-vacancy(NV) centre spin, which is generally considered as a promising candidate for similar experiments~\cite{scala2013matter,yin2013large,hammerer2015optomechanics,neukirch2015multi}. Another example is a rare-earth dopant spin in a crystal~\cite{siyushev2014coherent, liu2018controlled}.  The mass is initially optically trapped, made neutrally charged~\cite{frimmer2017} and rapidly cooled~\cite{deli2019motional,chang2010cavity,ferialdi2019optimal,rahman2017laser,mungan1999laser}. The internal spin state is then initialised by the application of a sudden microwave pulse in a superposition of spin eigenstates $\frac{1}{\sqrt{2}}\left(\left|+1\right\rangle + \left|0\right\rangle\right)$. At this point ($t=0$) the mass is released from the trap in a motional wavepacket $|\psi(0)\rangle$ centred at velocity $\bi{v}=\left(0,v_y,0\right)$ with the aforementioned internal spin superposition. The presence of a magnetic field gradient ($\partial_{x}\bi{B}$) in the $x$ direction induces an acceleration $\bi{a}=\left(a,0,0\right)$ on the $|+ 1\rangle$ spin state. The magnetic field gradient source we consider here consists of many flat carbon nanotubes arranged as shown in the detailed cut-out of Figure~\ref{fig:Interferometer path diagram}. To ensure a uniform magnetic field gradient is achieved the current through the wire can be switched on only when it is directly above the particle. This acts to generate the spatial superposition required while also coupling the spin and motional states.  The acceleration of the non-zero spin component is reversed at time $t=\tau_1$ and again at $t=\tau_2=3\tau_1$ by reversing the spin state, while the acceleration magnitude is maintained so that at any time $t$, the combined spin and motional state is $\frac{1}{\sqrt{2}}\left(\left|0\right\rangle |\psi_{0}(t)\rangle + \left|\sigma\right\rangle |\psi_{\sigma}(t)\rangle\right)$, where $\sigma$ represents the non-zero spin state. This procedure will lead to the maximum spatial superposition distance $\Delta x$ occurring at time $t=2\tau_1$, at which point the centres of the spatial states $|\psi_{0}(2\tau_1)\rangle$ and $|\psi_{\sigma}(2\tau_1)\rangle$ are separated by $\Delta x=a\tau_1^2$. These are then brought back together so that their motional states exactly overlap at time $t=\tau_3=4\tau_1$, i.e., $|\psi_{0}(4\tau_1)\rangle = |\psi_{\sigma}(4\tau_1)\rangle$. 

This spin-motion coupled interferometry has two striking consequences \cite{wan2016free}: (i) The relative phase $\Delta \phi$ between the interferometric arms is mapped on to the spin state in the form $\frac{1}{2}e^{i\phi_{0}(t)}\left(\left(e^{i\Delta\phi_{S}}-1\right)\left|0\right\rangle+\left(e^{i\Delta\phi_{S}}+1\right)\left|\uparrow\right\rangle\right)$, so that it can be measured by measuring the spin state alone. For example, by measuring the probability of the state to be brought to the spin state $|0\rangle$ after the application of a third microwave pulse. (ii) The $\Delta \phi$ depends solely on the difference between phases accumulated in the interferometric paths, and is quite independent of $|\psi(0)\rangle$ making the interferometric signal unaffected by an initial mixed thermal state or other noise (e.g. seismic) which occurs prior to the initialising microwave pulse, which can always be modelled as probabilistic choices of $|\psi(0)\rangle$. Any phase difference $\Delta\phi\ge\frac{1}{\sqrt{N}}$ will then be detectable after $N$ measurements.

Thus the whole interferometric process will lead to the state of the particles state evolving approximately as
\begin{eqnarray}
	&\textrm{Initial state:}&\left|0\right\rangle\otimes\left|\psi(0)\right\rangle \nonumber\\
	& \textrm{Microwave pulse:}& \frac{1}{\sqrt{2}}\left(\left|0\right\rangle+\left|\sigma\right\rangle\right)\left|\psi(0)\right\rangle \nonumber \\
	&\textrm{Spatial superposition created }& \nonumber\\
	&\textrm{and maintained for time $t$:}& \frac{1}{\sqrt{2}}\left(e^{i\phi_{0}(t)}\left|0\right\rangle\left|\psi_{0}(t)\right\rangle+e^{i\phi_{\sigma}(t)}\left|\sigma\right\rangle\left|\psi_{\sigma}(t)\right\rangle\right) \nonumber \\
	&\textrm{Spatial wavefunctions}& \nonumber\\
	&\textrm{brought to overlap:}& \frac{1}{\sqrt{2}}\left(e^{i\phi_{0}(t)}\left|0\right\rangle+e^{i\phi_{\sigma}(t)}\left|\sigma\right\rangle\right)\left|\psi(t)\right\rangle \nonumber \\
	&\textrm{Microwave pulse:}& \frac{1}{2}\left(\left(e^{i\phi_{\sigma}(t)}-e^{i\phi_{0}(t)}\right)\left|0\right\rangle+\left(e^{i\phi_{\sigma}(t)}+e^{i\phi_{0}(t)}\right)\left|\sigma\right\rangle\right)\left|\psi(t)\right\rangle \nonumber \\
	&\textrm{Final state:}& \frac{1}{2}e^{i\phi_{0}(t)}\left(\left(e^{i\Delta\phi_{S}}-1\right)\left|0\right\rangle+\left(e^{i\Delta\phi_{S}}+1\right)\left|\sigma\right\rangle\right)\left|\psi(t)\right\rangle \label{eq:single interferometer general output state}
\end{eqnarray}
Here $\left|\psi(t)\right\rangle$ is the original spatial state of the particle if it were to freely evolve and evaluated at time $t$, $\left|\psi_{0}(t)\right\rangle$ and $\left|\psi_{\sigma}(t)\right\rangle$ are the mass state in the spin-zero and non-zero arms of the interferometer respectively and $\left|0\right\rangle$ and $\left|\sigma\right\rangle$ are the respective spin states. This is an approximation of the evolution undertaken by the particle, whereby each effect is taken to occur stepwise. The magnetic field gradient state creates and recombines the spatial superposition, the microwave pulses create and recombines the spin superpositions. Of particular note is that the initial state of the mass factors in the final result, this will trivially hold in general, even if more complex states, for example thermal states, are used as the initial state.

This interferometric system amounts to an asymmetric modification of that proposed by Wan et. al.~\cite{wan2016free}. For a more in depth discussion of the required parameters required to realise the most sensitive and ambitious form of the interferometer we will propose can be seen in Section~\ref{sec:Practical Impelementation}.

\section{Observable Components of Space-Time Metric \label{sec:Observability of Space-time}}

To determine which components of the metric perturbation $h_{\mu\nu}$ are observable, we expand the action, $S$, to the second order in derivatives of $h_{\mu\nu}$ assuming a temporally static and spatially slowly varying metric. Specifically we take
\begin{eqnarray}
	h_{\mu\nu}(x,y,t)&\approx&h_{\mu\nu}(0,0,0)+x(t)\partial_xh_{\mu\nu}(0,0,0) + y(t)\partial_yh_{\mu\nu}(0,0,0) + \frac{1}{2!}\bigg({x(t)}^2\partial_x^2h_{\mu\nu}(0,0,0) \nonumber\\
	& & + {y(t)}^2\partial_y^2h_{\mu\nu}(0,0,0) + 2x(t)y(t)\partial_x\partial_yh_{\mu\nu}(0,0,0)\bigg)
\end{eqnarray}
For clarity  we will from now write $h_{\mu\nu}(0,0,0)$ as $h_{\mu\nu}$. This gives the difference in the action between the two interferometric paths due to the different components $h_{\mu\nu}$ ($\mu,\nu=0,x,y,z$) as
	\begin{eqnarray}
	\Delta S\left(h_{00}\right) &=& mc^2a \tau_1^3\left(\partial_x h_{00}+ \frac{23}{60}a\tau_1^2 \partial_x\partial_x h_{00} +  2v_y\tau_1\partial_x\partial_yh_{00}\right) \nonumber\\
	&=& mc^2a \tau_1^3\left(\partial_x h_{00}+ \frac{23}{60}a\tau_1^2 \partial_x\partial_x h_{00}+...\right),\label{eq:nwGeneralDeltaS}\\
	\sum_{j}\Delta S\left(h_{0j}\right) &=& mcav_y\bigg(
	2\tau_1^3 \left(\partial_x h_{0y}-\partial_y h_{0x}\right) + 4v_y\tau_1^4\left(\partial_y\partial_yh_{0x}-\partial_x\partial_yh_{0y}\right) \nonumber\\
	& &+ \frac{23}{30}a\tau_1^5\left(\partial_x\partial_x h_{0y}-\partial_x\partial_y h_{0x}\right)\bigg) \nonumber\\
	 &=& mcav_y\left(
	-2\tau_1^3 \partial_y h_{0x}+ 2\tau_1^3\partial_x h_{0y} + \frac{23}{30}a\tau_1^5\partial_x\partial_x h_{0y}+...\right)\,,\label{eq:fdGeneralDeltaS}\\
	\Delta S\left(h_{xx}\right)&=& -\frac{2}{3}ma^2\tau_1^3\bigg(h_{xx} + 2v_y\tau_1\partial_yh_{xx} + \frac{1}{2}a\tau_1^2\partial_xh_{xx} \nonumber\\
	& & +\frac{51}{20}v_y^2\tau_1^2\partial_y^2h_{xx}+\frac{43}{280}a^2\tau_1^4\partial_x^2h_{xx}\bigg) \nonumber\\
	\Delta S\left(h_{xy}\right)&=& mav_y^2\tau_1^3\partial_yh_{xy}+2mav_y^3\tau_1^4\partial_y^2h_{xy}+\frac{293}{60}ma^2v_y^2\tau_1^5h_{xy} \nonumber\\
	\Delta S\left(h_{yy}\right)&=& -mav_y^2\tau_1^3\partial_xh_{yy}-\frac{38}{3}mav_y^3\tau_1^4\partial_x\partial_yh_{yy}-\frac{23}{60}ma^2v_y^2\tau_1^5\partial_x^2h_{yy} \nonumber\\
	\sum_{i,j}\Delta S\left(h_{ij}\right)&=& \frac{-2}{3}h_{xx}ma^2\tau_1^3 + ...=\frac{-2}{3}h_{xx}mv_x^2\tau_1+...\label{eq:gwGeneralDeltaS}
	\end{eqnarray}		
The equations presented here are split such that, once truncated, they correspond to the Newtonian potential (Equation~\ref{eq:nwGeneralDeltaS}), Frame Dragging (Equation~\ref{eq:fdGeneralDeltaS}) and Gravitational waves (Equation~\ref{eq:gwGeneralDeltaS}) effects. Here we note that the example interferometer can directly detect certain components of the metric perturbation. Specifically the term $h_{xx}$ and, as rotating the apparatus is equivalent to relabelling the spatial direction, the spatial components of the metric in general. Furthermore as the action is directly dependent on the second derivatives of $h_{\mu \nu}$, such an apparatus would also be sensitive to the local space-time curvature \footnote{The complete curvature is characterised by the Riemann tensor which is defined by $R_{\mu\nu\sigma\nu}=\frac{1}{2}\left(\partial_{\sigma}\partial_{\mu}h_{\rho\nu} + \partial_{\nu}\partial_{\rho}h_{\mu\sigma} - \partial_{\nu}\partial_{\mu}h_{\rho\sigma} -\partial_{\sigma}\partial_{\rho}h_{\mu\nu}\right) \label{eq:Riemann tensor}$.}. This allows the experimentalist to simply identify certain components of the Riemann tensor $R_{\mu\nu\sigma\nu}$ in the above equations term by term, it is for this reason we consider the interferometer directly sensitive to space-time curvature. The role of the asymmetry in the interferometer can also now be seen from \ref{eq:gwGeneralDeltaS}, given the second order terms $a^2\tau_1^2\approxeq\left(v_x\right)^2$ dependence, asymmetry is necessary to generate an action difference between the arms. For example, if a symmetric interferometer was used, by taking the initial spin state of $\frac{1}{\sqrt{2}}\left(\left|+1\right\rangle+\left|-1\right\rangle\right)$, then both arms would contain the same $v_x^2$ dependent phase as seen in Equation~\ref{eq:gwGeneralDeltaS}. These would cancel in the final phase difference, leaving the interferometer no longer sensitive to GWs.

In the following sections we will explore the basic experimental considerations for detecting Newtonian gravity and its associated curvature (Section~\ref{sec:Newtonian potential}), frame dragging effects (Section~\ref{sec:Frame dragging}) and gravitational waves (Section~\ref{sec:Gravitational Waves}). We present the exact form that the signals will take, discuss their predicted amplitudes to discern how well they can be detected and for the case of the Newtonian potential, we will also explore and characterise a variety of different sources which could be generating the signal. 

\section{Newtonian potential \label{sec:Newtonian potential}}	
\begin{figure}
		\centering
		\includegraphics[width=\linewidth]{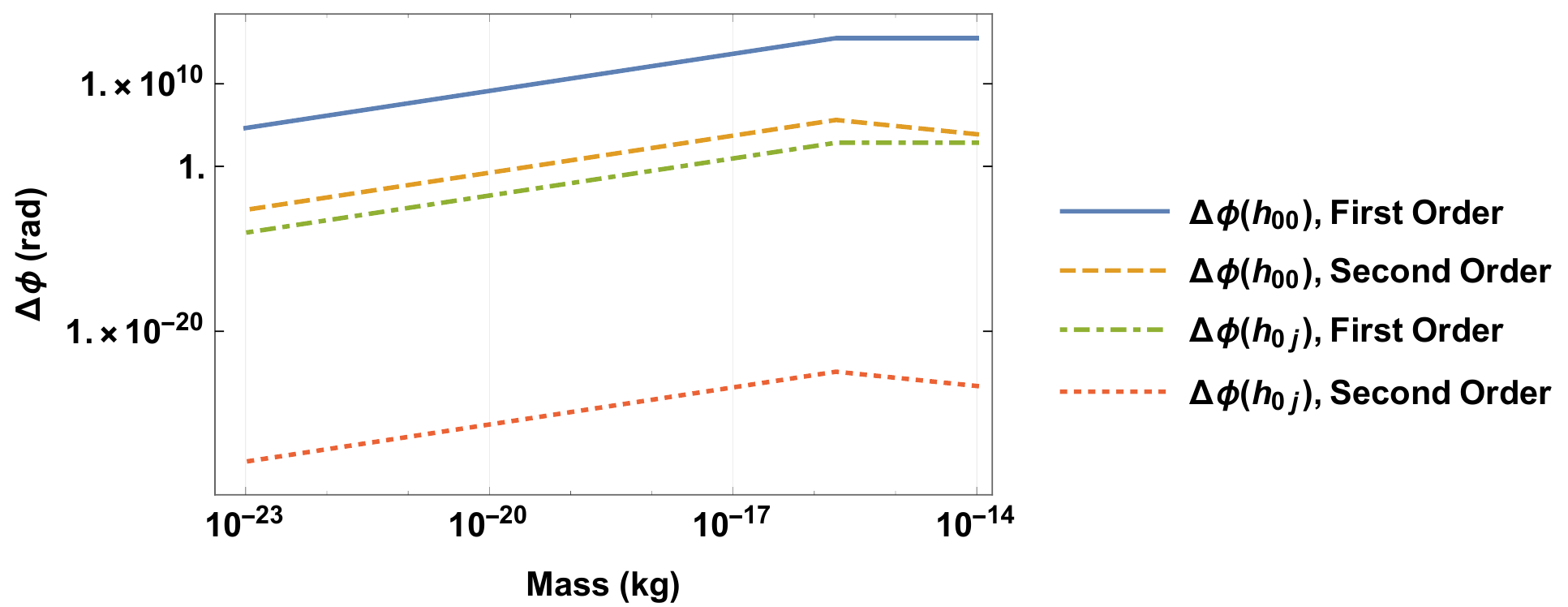}
		\caption{Newtonian potential and frame dragging phase difference scaling with the mass of objects for a maximum interferometer size and time of $\Delta x = 1\textrm{ mm}$ and $\tau_1 =100$ ms respectively with $v_y=10\textrm{ ms}^{-1}$. As the mass $m$ increases, the phase change increases as $\Delta x=a\tau_1^2$ can be kept to its highest value by allowing more time $\tau_1$. However, an optimal point is reached slightly after about $m=10^{-16}$ kg after which the $\Delta x$ obtained with the maximum $\tau_1$ starts decreasing in inverse proportion to mass even for the fixed maximum feasible values of magnetic fields ($10^{6}\textrm{ Tm}^{-1}$). \label{fig:DeltaPhis}}
	\end{figure}

Considering only the first non-Minkowski term in \ref{eq:General action formula} we can make the standard substitution for the Newtonian potential, $h_{00}= 2MG/c^2R$. We define the vertical as the $x$-axis, the experiment taken to be performed at ground level so $R$ is radius of the Earth, and $M$ Earth's mass, the difference in action between the two arms up to the second order in $\left(a\tau_1^2/R\right)$ is
	\begin{eqnarray}
		\Delta S\left(h_{00}\right) &\approx& -\frac{2mMG}{R^2}a\tau_1^3+\frac{23mMG}{15R^3}a^2\tau_1^5 \,,\label{eq:Delta S Newtonian} 
		\\\Delta\phi\left(h_{00}\right) &\approx& -2\times 10^{35} \textrm{ kg}^{-1}\textrm{m}^{-1}\textrm{s}^{-1} \times ma \tau_1^3 \nonumber\\
		& & + 2\times 10^{28} \textrm{ kg}^{-1}\textrm{m}^{-2}\textrm{s}^{-1} \times ma^2\tau_1^5.  \label{eq:Newtonain phase without magnetic field gradient}
	\end{eqnarray}
This is consistent with the notion that any curvature detection will be of the form $U ({L}/{R})^2$ where $U$ is the gravitational potential and $L$ is the characteristic laboratory length (in the above case, $L\sim a\tau_1^2$)~\cite{Visser2018}. Despite this quadratic suppression, it is still detectable due to the $1/\hbar$ factor in the phase difference. As such, we can expect to observe even second order effects (curvature effects) as large phase shifts. Figure~\ref{fig:DeltaPhis} shows how these results scale with the mass of the object in the interferometer assuming a maximum allowed value of the spatial separation ($a\tau_1^2$).
From figure~\ref{fig:DeltaPhis} it can be seen that a mass of $10^{-16}\textrm{ kg}$ in a $\sim1\textrm{ mm}$ interferometer with integration time $\tau_1\sim100\textrm{ ms}$ gives a detection of acceleration with sensitivity down to $\sim 5\times10^{-15}\textrm{ ms}^{-2}\textrm{Hz}^{-1/2}$. This result is for the case of sending a single particle through the interferometer at a time and as such represents a lower bound on the sensitivity of such a detector. This compares favourably with the recent work demonstrating the direct detection of metric curvature of a test mass with a sensitivity of $~5\times10^{-9}\textrm{ ms}^{-2}\textrm{Hz}^{-1/2}$~\cite{asenbaum2017phase}

This detector could also be used to detect smaller masses and more local signals. For example, the mass $M$ at distance $R$ which yields a detectable phase shift compared to it not being there, effectively it ceasing to exist, is given by
\begin{equation}
	M=\frac{\hbar R^2}{2\sqrt{N}mG\Delta x \tau_1} \label{eq:Mass detectable}
\end{equation}
which suggests for the interferometer specifications used for figure~\ref{fig:DeltaPhis}, at a distance of $1$ km, a mass of approximately $4$ kg is detectable provided the mass has moved from a very far distance to this $1$ km range or by varying the interferometer orientation relative to the mass. On the other hand, all stationary masses naturally present around the interferometer will not act as a noise when detecting other signals as they will provide a constant phase difference between arms for a fixed orientation of the interferometer.

We can also consider detecting the motion of a mass. Taking the motion to be slow enough that the interferometer phase can be found for the mass $M$ at $R$ before it moves a distance $d$ and detected again. The minimum movement detectable will then be
\begin{equation}
	d\approx\frac{\hbar R^3}{4\sqrt{N}mMG\Delta x \tau_1} \label{eq:Min movement detectable}
\end{equation}

where it has been assumed that $d\ll R$. For example the previous $M=4$ kg mass a distance $R=100$ m away will produce a detectable phase variation if it moves by $d\approx0.5$ m or more. This can act as a noise source when looking to detect other signals, this will be discussed below in Sec.~\ref{sec:Practical Impelementation}.

\section{Frame Dragging \label{sec:Frame dragging}}
 To explore the detection of frame dragging, the `frame dragging' metric given in~\cite{Weinberg1972} was considered. Written in  spherical with $\psi$ the azimuthal and $\theta$ the polar angles:
\begin{equation}
ds^2=-H\left(r\right)~c^2dt^2 + J\left(r\right)\biggl[dr^2 + r^2~d\theta^2 + r^2\sin^2\left(\theta\right)\left(d\psi-\Omega~dt\right)^2\biggr] \label{eq:Full frame dragging metric line element}
\end{equation}
where
\begin{equation}
H\left(r\right) \approx 1-\frac{8GM}{c^2r}+\cdots\,,~ J\left(r\right) \approx 1+\frac{8MG}{c^2r} +  \cdots \,, \label{eq:Frame dragging J(r) with correct units}
\end{equation}	
where the binomial expansion approximation has been used for being in the linearized limit, and  $\Omega={2MG\nu}/{c^2R}$ is the scaled angular velocity of the central rotating mass, where once again $M$ is the mass of the Earth, $R$ is its radius and $\nu$ is its angular velocity. The relevant component of \ref{eq:Full frame dragging metric line element} is the cross term $d\psi dt$.

The apparatus is taken to be aligned parallel with the equator and surface of the earth, and taking a small angle approximation with regards to the angular distance the mass travels along the interferometer in the `$y$' direction measured from the centre of the earth. Defining $M$ as the mass of the Earth, $R$ its radius, and $\nu$ its angular velocity gives a phase difference, again to the second order in $\left(a\tau_1^2/R\right)$
	
\begin{equation}
	\fl \Delta\phi\left(h_{0j}\right)\approx \frac{8mMG\nu\sin^2\left(\theta\right)av_y}{\hbar c^2 R}\biggl(\tau_1^3 - \frac{3M^2G^2}{c^4 R^2}\tau_1^3\biggr) +\frac{92mM^3G^3\nu\sin^2\left(\theta\right)v_y}{5\hbar c^6 R^4}a^2 \tau_1^5. \label{eq:Delta Phi Frame Dragging}
\end{equation}

Substituting all known constants, assuming the interferometer is located on the surface of the Earth, gives $\Delta\phi\left(h_{0j}\right) \approx4\times10^{21}mav_y\tau_1^3$ as the first order, metric dependent phase and $\Delta\phi\left(h_{0j}\right) \approx6\times10^{-4} ma^2v_y\tau_1^5$ for the second order, curvature dependent phase. These effects are significantly more modest so high precision measurements would be needed, specifically to measure the second order term. Such measurements would provide an independent verification of the results from Gravity Probe B~\cite{GravityProbeB}. Figure~\ref{fig:DeltaPhis}  also shows the phase due to first and second order effects independently with respect to the object mass. 

\section{Gravitational waves \label{sec:Gravitational Waves}}	
	Our setup can also extract the phase from the transverse traceless perturbations around the Minkowski background:
	
	\begin{eqnarray}
		h_{xx}=-h_{yy}&=h_+\cos\left(\psi_0+\omega t\right) \\
		h_{xy}=h_{yx}&=h_{\times}\cos\left(\psi_0+\omega t\right). 
	\end{eqnarray}
where $\psi_0$ is the GW phase at the interferometer at $t=0$ in the interferometers reference frame. We have assumed the GW is propagating along the $x_{3}=z$ direction perpendicularly to the interferometer with angular frequency $\omega$ and taken the two helicity states of the GWs as $h_+, ~h_\times\ll 1$. We also ignore the kinetic energy component of the atoms action, see \ref{eq:General action formula}, as it is not relevant for the purpose of detecting the phase. The GW induced phase difference is
	
	\begin{eqnarray}
		\Delta\phi\left(h_{ij}\right) &=& \frac{4mh_+a^2\tau_1\cos\left(\psi_0\right)\cos\left(\omega\tau_1\right)}{\hbar\omega^2}\left(1 -\frac{\sin\left(\omega\tau_1\right)}{\omega\tau_1}\right) \label{eq:Delta phi gravity wave 1}\\
		&\approx& \frac{2mh_{+}a^2\tau_1^3\cos\left(\psi_0\right)}{3\hbar} \label{eq:Delta phi gravity wave approximation}		
	\end{eqnarray}
where $\psi_0$ is the wave's phase at $t=2\tau_1$ and the approximate form holds when $\omega\tau_1\ll1$. Note the $h_{\times}$ component is not recorded in our interferometer, as it is proportional to $v_x$ which varies between  positive and negative values, thus cancelling itself out unlike $h_{+}$ as it is a function of $v_x^2$. A rotated apparatus detects $h_{\times}$.

The underlying mechanism for this phase difference is ultimately through the particle coupling to the local space-time parameters (the metric). The metric is what will be directly affected by the GW and this is detected through the phase evolution as given by the action, see \ref{eq:General action formula}. Note that our apparatus is \emph{not} directly detecting the tidal acceleration of the particle caused by the GW. In fact, it is negligible compared to that generated by the magnetic field gradient needed to enable the interferometry. It is simply measuring the spatial stretching/ contraction as caused by the GW in the same manner as it would measure a permanent change in the relevant components of the metric. Of course there is an unavoidable time variation of the metric due to the GW, but we do not exploit this variation\footnote{Here we are specifically referring to the variation during a single particles traversal of the interferometer. The sinusoidal modulation of the phase difference due to the variation of the metric from one run to the next will be how the GW is measured} -- the time variation of the metric is much slower than an individual run of the interferometer for the frequencies our detector is most sensitive to. Essentially the interferometer detects phase changes due to static metric components. In this way the correct analogy here between laser interferometers and our interferometer is that the mass is the replacement of the photons. They both act to measure the change in spatial distances due to the GW. As the path length difference of $\sim h_{+} L$ is essentially being measured in units of the matter wave de Broglie wavelength, $\sim 10^{-17}$ m, $L \sim 1$ m suffices (note in our case $L=\Delta x$). Let us emphasize here that one should {\emph{not}} interpret our interferometer as detecting the tidal acceleration as given by $h_{+}L\omega^2$ directly acting on the mass. This also leads directly to how the GW sensitivity in our interferometer scales uniquely compared to the acceleration sensitivity. Consider increasing the magnetic field gradient applied, such that $a\tau_1^2=\Delta x$ remains fixed, while reducing $\tau_1$. The GW induced phase difference scales as $\Delta \phi(h_{ij})\propto \frac{\Delta x^2}{\tau_1}h_{ij}$ because the GW metric couples to the velocity of the particle ($S\propto h_{ij}v^{i}v^{j}$) while the stray acceleration induced phase difference scales as $\Delta \phi\left(h_{00}\right)\propto\Delta x \tau_1h_{00}$. As such the GW sensitivity can be further enhanced while suppressing the noise effects in our signal, giving an improved signal to noise ratio. Thus our interferometer is qualitatively very different from LIGO/LISA. A second \emph{crucial} difference between laser interferometers and MIMAC is that there is no back-action and as such the related standard quantum limit is not a limiting factor. This is because the measurement only occurs \emph{once} after the interferometry has taken place, and the position is not measured either, only the final spin state. Indeed our interferometer is closest in mechanism to single atom interferometers, which were suggested as some of the early atom-interferometry schemes for GW detection \cite{Chiao2004,Roura2006,Foffa2006}. 

These two differences form the basis of the potential future advantages this interferometer holds over laser interferometers, in which the standard quantum limit and Newtonian noise act as the primary limits on the sensitivity. Neither are fundamentally limiting with MIMAC or a MIMAC like interferometer. 

\begin{figure}
	\centering
	\includegraphics[width=\linewidth]{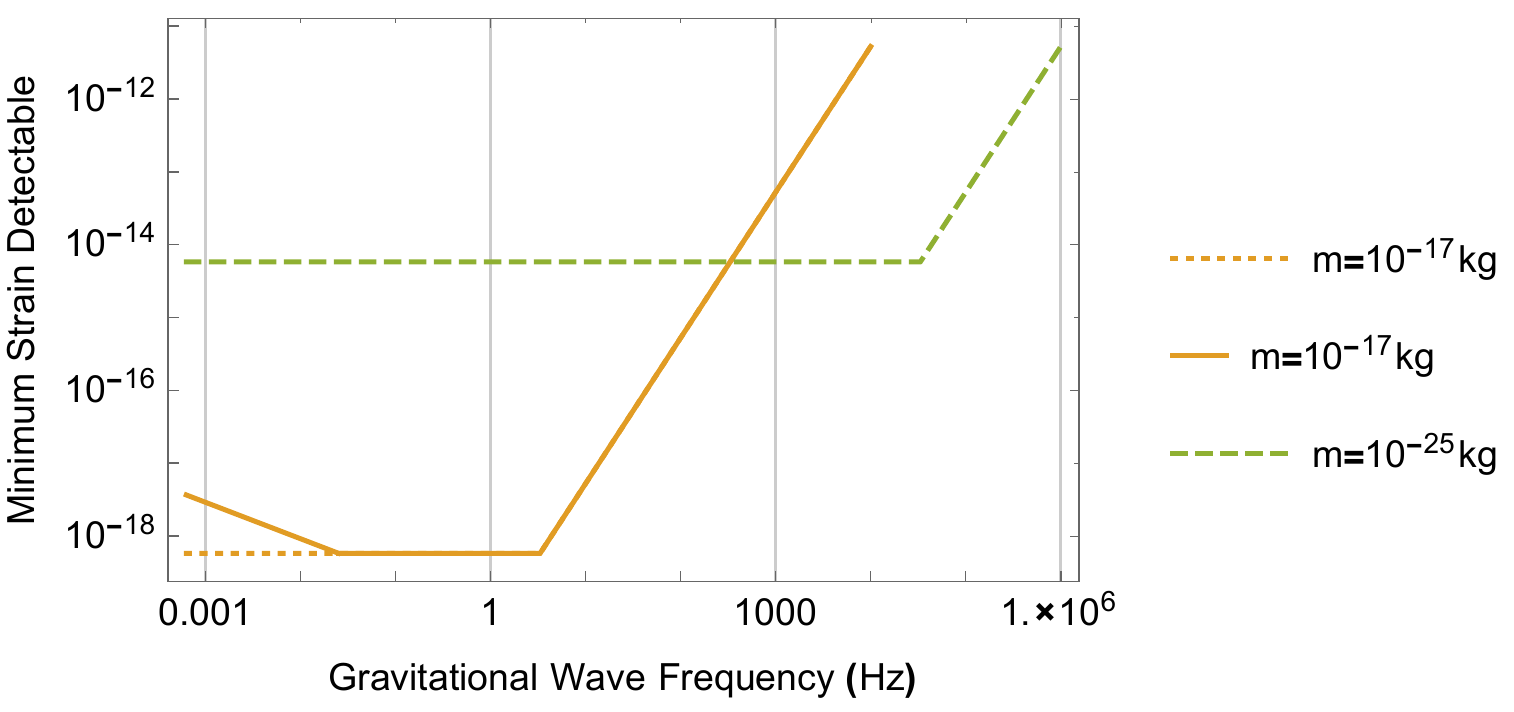}
	\caption{Comparison of strain sensitivity between two different mass, $\Delta x=1$ m GW detectors. The dashed green curve is for a ground based interferometer of mass $10^{-25}\textrm{ kg}$, $\tau_1=7.3\times10^{-5}\textrm{ s}$ and a flux $N=10^6$ taken from~\cite{Dimopoulos2008a} for $\textrm{Rb}^{87}$ atoms, the lack of GGN limit in the sensitivity can be attributed to the extremely short interferometry time reducing the effect of the Newtonian potential in the final phase difference. The lower orange curves are for a $10^{-17}\textrm{ kg}$ mass, $\tau_1\approx0.73\textrm{ s}$, and $N=400$ for ground based (solid, including GGN with relevant cancellation) and space based (dotted) sensitivities. It also shows the low frequency strain sensitivity reduction due to gravity gradient noise. \label{fig:GWStrainComparison}}
\end{figure}

With respect to the early atom interferometers, our advantage stems from the much larger $m$ for our interferometers as our SG methodology opens up the scope to create a high enough $\Delta x$, even with the increased mass. Here we should note that the more advanced proposals from atom interferometry such as Atomic GW Interferometric Sensor (AGIS) as discussed in~\cite{Dimopoulos2009} are qualitatively very different from our scheme. As such, we can compare only the scales, but not the mechanism. They generate a phase difference $\sim10^{16}h_{+}$ for the space based detector~\cite{Dimopoulos2008} with baseline size $L\sim10^{7}$ m compared to our $\Delta\phi\left(h_{ij}\right)\sim10^{17}h_{+}$ for a baseline size of $1$ m as shown in figure~\ref{fig:GWStrainComparison}. Again, as the mechanism of our proposal differs significantly from AGIS and related schemes the above comparison does not capture the entire effectiveness of these two proposals. 
	
One can see from \ref{eq:Delta phi gravity wave approximation} that the phase output will be independent of GW frequency provided $\omega\tau_3 \sim \omega\tau_1 \ll 1$, though it will be limited by gravity gradient noise at lower frequencies (see figure~\ref{fig:GWStrainComparison}). It is in this regime that our interferometer is most sensitive to GWs. The frequency scaling of detectability is understood by noting it is susceptible to the wave's time-averaged amplitude, which tends to zero for higher frequencies. As such, higher frequency GWs can be detected by using shorter time detectors, as seen in figure~\ref{fig:GWStrainComparison}, albeit with a lower sensitivity without also increasing the magnetic field gradient and mass. Note that we define a detectable strain by $\Delta\phi\left(h_{ij}\right)\ge1/\sqrt{N}$ for $N$ particles traversing the interferometer in series (and/or several interferometers in parallel). Further note that around $10-10^{4}$ Hz, LIGO is already performing\cite{Martynov:2016fzi}, while there are undetected lower frequency GW sources \cite{ellis2018maximal}. Our interferometer will be complementary in part of the range of LISA~\cite{LISA} ($10^{-6}-10$ Hz) for an underground implementation or all of its range for a space based interferometer.
	
\section{Practical Implementation \label{sec:Practical Impelementation}}

While the sensor we have proposed is ambitious in its scope, there does not appear to be any fundamental or insurmountable obstacle to its creation using current and near future technologies. Furthermore, we are primarily looking to show its `in principle' feasibility by presenting an example scheme for realising the interferometer. For the remainder of this article we will outline the techniques which can be employed to create such an interferometer. We will discuss the primary sources of decoherence which act to destroy the superposition as well as consider the primary sources of noises in the phase output signal. This will be used to put limits on the tolerable noise and fluctuations of the experimental parameters such as mass fluctuations from one particle to the next and timings. On top of the constraints and methods discussed below, the creation of this interferometer will require further work to ensure excellent surface termination to reduce dangling bonds, motional decoupling and a method for the creation of a beam of flying diamond among further experimental advances on which work is ongoing~\cite{pedernales2019motional} in the relatively new field of large mass interferometry.

To realise the proposed interferometer a magnetic field gradient ($\partial_{x}\bi{B}$) is used to create the spatial superposition of size $\Delta x =a\tau_1^2$ with $a=g_{NV}\mu_{\textrm{B}}\partial_x\bi{B}/m$ where  $g_{NV}$ is the Land\'e \textit{g} factor and $\mu_{\textrm{B}}$ is the Bohr magneton~\cite{wan2016free}. For large mass interferometry to carry advantage over atoms, $\Delta x$ must be kept $\sim1$m even while $m$ increases. To this end, if we are to keep $\tau_1\approx0.73$ s as is required to achieve our maximum  GW sensitivity (see figure \ref{fig:GWStrainComparison}) a magnetic field gradient, $\partial_{x}\bi{B}$ of $10^6$ Tm$^{-1}$ is needed. Such a large magnetic field gradient could be created using a current carrying wire. We however propose the use of dual overhead wires. This allows for a more uniform magnetic field gradient to be maintained while increasing the distance between the interferometric particles and the wires, so reducing spurious forces. These wires would have to arranged in many small horizontal sections such that they approximately follow the path of the non-zero spin interferometer arm, as shown in Figure~\ref{fig:Interferometer path diagram}. This allows it to always remain proximal to the non-zero spin interferometer arm, generating a sufficiently large magnetic field gradient without also requiring an unreasonably large magnetic field. This requires a large current, which will necessitate the use of carbon nanotube-metal composites, which can support a current density of up to $\rho_I=10^{13}$ Am$^{-2}$\cite{subramaniam2013one}. The magnetic field gradient amplitude from a single wire is
	
\begin{eqnarray}
	B&=\frac{\mu_0 I}{2 \pi D}& \nonumber\\
	\partial_{x} B&=\frac{\mu_0 I}{2\pi D^2} &= \frac{\mu_0 \rho_I \tilde{r}^2}{2 \left(\tilde{r}+\Lambda\right)^2} \nonumber\\
	&\approx\frac{\mu_0 \rho_I}{8} &\sim 10^{6} \textrm{~Tm}^{-1}
\end{eqnarray}
where here $D$ is the distance between the centre of the wire and the point at which the magnetic field strength is measured, $\tilde{r}$ is the radius of the wire, $\Lambda$ is the distance from the surface of the wire and we have taken ($\Lambda=\tilde{r}$). In this way, the primary concern to creating the large magnetic field gradients necessary are the current stability and the distance $\Lambda$ required to eliminate other interactions, such as the patch potential and Casimir interactions, importantly this distance simply sets the thickness required, and does not limit the theoretical possibility of achieving the required magnetic field gradient. To generate a sufficiently uniform magnetic field gradient we propose many small pairs of overhead wires are used which modifies the experienced magnetic field gradient slightly. However, for clarity and as this is a simple proof of concept argument being presented, a simpler (single bent wire) set-up will be discussed in detail below as numbers wise, the gradient strength, noise and decoherence are effectively the same.

\subsection{Decoherence \label{subsec:Decoherence}}
The primary sources of decoherence for the spatial superposition states will be scattering of air molecules and black-body emission giving ``which path'' information. The spatial coherence can offer a huge window using low pressure $\sim10^{-14}\textrm{ Pa}$ (with lower achieved previously in cryogenically cooled systems \cite{Gabrielse2016}) and low internal temperature $\sim50$ mK. This is  achievable, for example, in a dilution fridge~\cite{bose2018matter} or using laser cooling~\cite{rahman2017laser,mungan1999laser}. For a mass of $\sim 10^{-17}$ kg and $100$ nm radius, using the results of \cite{Romero-Isart2011}, scattering rates are calculated to be $0.006\textrm{ Hz}$ due to scattering of air molecules and $0.06\textrm{ Hz}$ due to black-body photon emission. The electron spin coherence at $10$ mK can also reach $1$ s with dynamical decoupling \cite{Bar-Gill(2013),Abobeih(2018)} (partially present here due to spin flipping pulses, and further extendable by applying pulses to the spin bath \cite{knowles2014observing}. The scale of the superpositions considered here are consistent with stochastic GW induced decoherence as we use mesoscopic objects \cite{steane2017matter}.

The proposed set-up brings together state of the art magnetic field gradient, pressure and internal temperature, all having been realized individually. Sending  nanodiamonds through low pressure is still being developed, as well as combining free flight with cryogenics \cite{bose2018matter}.

\subsection{Gravitational Signals as Noise \label{subsec:Gravitational signals as noise}}

By construction, our interferometric signal only depends on the relative phase between the two arms and thereby is immune to thermal and seismic noise in $|\psi(0)\rangle$.  Thus for the most sensitive proposed interferometer (for GWs), the phases due to frame dragging and Newtonian potential type sources (including gravity gradients \cite{hughes1998seismic}) are the primary noise. There can also be further noise sources due to the implementation, for example particle-particle and particle-magnet casimir, patch potential and gravitational interactions. 

In the following, we will consider the most challenging to detect signals (GWs) for which the highest strain sensitivity of $h_+ \sim 10^{-17}$ occurs for \emph{single} masses of $\sim 10^{-17}$ kg, each traversing the interferometer one at a time. We can stretch this to $h_+ \sim 10^{-18}$ by considering $N=400$ masses traversing the interferometer in series over the duration of the interferometer ($\tau_3$), one after another. This can be achieved by successively cooling~\cite{deli2019motional,chang2010cavity,ferialdi2019optimal,rahman2017laser,mungan1999laser} and injecting one particle every $\sim10$ ms. This then sets the signal strength which all noises must be kept below (we will discuss below how this can be met). Further, for low frequency GW detection, say for GWs of frequency $\sim 10$ mHz one can do $\sim100$ repeats of this interference during the period of the gravitational wave. This will improve the sensitivity by an order of magnitude so as to bring the detector into the range of detection of massive binaries at the above frequency~\cite{moore2014gravitational}. One can further improve sensitivity by another factor of $1/\sqrt{\mathcal{N}}$ by using $\mathcal{N}$ interferometers in parallel. This also corresponds to the most ambitious setting for the sensor, with a mass $m=10^{-17}$ kg, time $\tau_1=0.73$ s, $v_x=a\tau_1=1.35$ ms$^{-1}$.

Firstly, a simple source of noise in any signal, GW signals included, is due to parameter fluctuations from one run to the next. With this in mind it is only necessary to consider the largest phase effect (the Newtonian potential) as it will magnify any uncertainty the most It should be noted that, although not immediately obvious from \ref{eq:nwGeneralDeltaS} or~\ref{eq:Delta S Newtonian}, the first order Newtonian phase is independent of the particles mass, this is due to the inverse scaling of the superposition size with the mass. Furthermore these noises can be suppressed by orientating the interferometer to be perpendicular to the Newtonian potential gradient (parallel with the ground). This gives a phase uncertainty $\delta\phi$ due to mass ($\delta m$), distance ($\delta R$), superposition size ($\delta(\Delta x)$) and timing ($\delta \tau_1$) uncertainty of approximately

\begin{equation}
\fl\delta\phi\approx\frac{23 \delta m MG\sin(\alpha)}{15\hbar R^3}\Delta x^2\tau_1 +\frac{2mMG\sin(\alpha)}{\hbar R^2}\left(\frac{-2\delta R}{R}\Delta x\tau_1 \nonumber + \delta(\Delta x)\tau_1 + \Delta x \delta\tau_1\right)
\end{equation}
where $\alpha=0$ when the interferometer is exactly perpendicular to the local Newtonian potential gradient. This was derived from Equation~\ref{eq:Delta S Newtonian} allowing for variations in the experimental parameters and orientating the interferometer relative to the local Newtonian gravitational potential. Given that an orientation uncertainty $\le1$ pRad is measurable~\cite{hogan2011precision}, thus $\left|\sin\left(\alpha\right)\right|\le10^{-12}$ is achievable, the mass, distance, separation and timing fluctuation would have to be kept below $\delta m\le10^{-18}$ kg, $\delta R\le0.1$ m $\delta(\Delta x)\le10$ nm and $\delta \tau_1\le1$ ns respectively to ensure $\delta\phi$ is kept below the detectable limit, that is, to ensure $\delta\phi\le0.1$. Variations in otherwise known (systematic) phases can be countered through a careful characterisation of system parameters and/or modifications of the interferometric setup.

We can note that some noises can be identified due to the unique functional dependences (specifically how they scale with $a$, $v_y$ and $\tau_1$) of the 5 identified signals (\ref{eq:nwGeneralDeltaS}-\ref{eq:gwGeneralDeltaS}) the individual types of signals could be identified separately by a network of interferometers allowing them the signal to be filtered out from them. Of specific note is that by setting $v_y=0$, \ref{eq:fdGeneralDeltaS} becomes zero. Doing so however will limit the ability to introduce more then one particle into the interferometer at a time, making the sensitivity (and noise ceiling) $\Delta\phi=1$ for a single run of a single interferometer. Furthermore, certain external noises can be actively cancelled.  First order signals can be detected and cancelled by a symmetric detector (using an initial spin superposition $\frac{1}{\sqrt{2}}\left[|+1\rangle+|-1\rangle\right]$) insensitive to second order effects and GWs. Here by first order signals we are referring to terms in \ref{eq:nwGeneralDeltaS} and~\ref{eq:fdGeneralDeltaS} which are a function of a single derivative. This can be done as these are the only terms which a symmetric interferometer is sensitive to (the $\propto a^2$ terms cancel when the difference is calculated), as such, these noises can be treated as signals which can be subtracted from the total phase output. The second order Newtonian potential term can also be approximated by the use of slightly displaced symmetric interferometers. These would again be insensitive to GWs and would result in third order effects being left in the noise. This method of active cancellation is also only an approximation. For example consider a source located a distance $R$ from the primary detector, with secondary, symmetric interferometers located at $R\pm s$ from the source. The signal at the central asymmetric interferometer would be approximately the average of signal at each symmetric interferometer either side of it. This approximate signal can be used to cancel the phase noise, thus reducing it by a factor $\epsilon^{\left(1\right)}$ which encompasses how close the approximation is. To determine $\epsilon^{\left(1\right)}$ we can expand the signal in orders of $\frac{s}{R}$ from the central, asymmetric interferometer giving

\begin{eqnarray}
\epsilon^{\left(1\right)}&=&\frac{\delta\phi^{\left(1\right)}\left(R\right)-\frac{1}{2}\left(\delta\phi^{\left(1\right)}\left(R+s\right)+\delta\phi^{\left(1\right)}\left(R-s\right)\right)}{\delta\phi^{\left(1\right)}\left(R\right)} \nonumber\\
&=&1-\frac{1}{2}\left(\left(1+\frac{s}{R}\right)^{1/2}+\left(1-\frac{s}{R}\right)^{1/2}\right) \nonumber\\		&\approx&1-\frac{1}{2}\left(1+\frac{s}{2R}-\frac{s^2}{8R^2}+1-\frac{s}{2R}-\frac{s^2}{8R^2}\right)\nonumber\\
&=&-\frac{s^2}{8R^2}.
\end{eqnarray}
Take for example the movement of a $1$ kg mass, a distance of $1$ m away from the sensor and aligned with the interferometers $x$ axis (the direction it is sensitive in). If we consider the primary interferometer as having a symmetric interferometer above and below it at a distance of $s=1$ cm then by \ref{eq:Min movement detectable} its movement would have to be less than $d=10^{-10}$ m without any active cancellation, however, with cancellation this becomes $d=10^{-5}$ m, a still significant, but far less difficult value.

\subsection{Gravity Gradient Noise \label{subsec:GGN}}
Distant Newtonian potential fluctuations are known as Gravity Gradient Noise (GGN)~\cite{PhysRevD.88.122003, Geiger2016}. This is known to be one of the primary noise sources which limit GW detections in present day GW antennas, particularly at the low frequencies. Gravity Gradient noise is due to seismic waves causing  variations in the local gravitational field. These seismic waves are not as dramatic as earthquakes, but stochastic fluctuations in the local density and surface fluctuations in the surrounding ground. It is difficult to say anything too specific about gravity gradient amplitudes as these are known to be highly location dependent~\cite{harms2009simulation}. We will be following closely the analysis performed in ~\cite{saulson1984terrestrial} and~\cite{pitkin2011gravitational}, combined with measured gravity gradient accelerations~\cite{harms2015terrestrial, nishida2017ambient} as well as consider how well we could hope to cancel such effects.

Consider the effect of a fluctuation in the atmospheric or ground density $\Delta\rho$ of some volume $V$, where for the example of ground based fluctuations of wavelength $\lambda$ and height $\xi$, $V=\lambda^2\xi$, at some distance $r$ from our interferometer. This will yield an anomalous acceleration of magnitude

\begin{equation}
a=\frac{G \Delta\rho V}{r^2}\cos\left(\beta\right)\sin\left(\gamma\right)
\end{equation}

where $\beta$ and $\gamma$ are the polar coordinates of the disturbance with the coordinate origin located at the detector. This was derived by considering the standard formula for acceleration due to the Newtonian gravitational interaction and that the interferometer is sensitive in only a single direction. Thus the trigonometric dependences are due to the directional sensitivity of the detector. To simplify the analysis we will consider all regions of fluctuation as independent and so consider the joint effect by adding the squared acceleration. We will also consider a minimum distance, $r_0$, that is our interferometer to be within a cavity in which there are no density fluctuations. Considering initially an interferometer located at the surface of the earth, then the square of the expected acceleration will be

\begin{eqnarray}
a^2&\approx& G^2 \Delta\rho^2V^2\int_{0}^{\frac{\pi}{2}}\int_{-\pi}^{\pi}\int_{r_0}^{\infty}\frac{1}{r^4}\cos^2\left(\beta\right)\sin^2\left(\gamma\right)r^2\sin\left(\gamma\right) \rmd r\rmd\beta \rmd\gamma \\
\rightarrow a &\approx& \sqrt{\frac{2\pi}{3}}\frac{G\Delta\rho V}{\sqrt{r_0}}
\end{eqnarray}
Now if the interferometer is placed underground at a depth $d$ this becomes
\begin{eqnarray}
a^2&\approx& G^2 \Delta\rho^2V^2\int_{0}^{\frac{\pi}{2}}\int_{-\pi}^{\pi}\int_{d/\cos\left(\gamma\right)}^{\infty}\frac{1}{r^2}\cos^2\left(\beta\right)\sin^3\left(\gamma\right)\rmd r\rmd\beta \rmd\gamma \\
\rightarrow a_{\textrm{u}}&\approx& 0.6\frac{a\sqrt{r_0}}{\sqrt{d}}.
\end{eqnarray}
and so we can estimate the phase noise in our underground sensor. Using the measured median results in~\cite{harms2015terrestrial, nishida2017ambient} of $a_{\textrm{surface}}=3\times10^{-11}$ ms$^{-2}$ for fluctuations occurring at $1$ mHz we can estimate the underground phase noise due to a stochastically varying local acceleration assuming $r_0=1$ m and $d=100$ m as

\begin{eqnarray}
\delta\phi^{\left(1\right)}&\sim& \frac{2m}{\hbar}\left(0.6\frac{a\sqrt{r_0}}{\sqrt{d}}\right)\Delta x\tau_1 \\
&\sim& 2\times 10^{5}
\end{eqnarray}
which is clearly quite significant. It is also worth noting that `quiet' (low GGN noise) sites can have noise values two orders of magnitude smaller~\cite{saulson1984terrestrial}. There will also be second order effects ($\delta\phi^{\left(2\right)}$) where the local gravity varies across the interferometer which will be approximately a factor of $\frac{\Delta x}{\lambda}\sim 0.001$ smaller for typical fluctuation wavelengths $\lambda=1$ km~\cite{harms2009simulation} giving $\delta\phi^{\left(2\right)}\sim2\times10^{2}$ at $1$ mHz. 

These can however be measured and cancelled using symmetric implementations of the interferometer as discussed above, hence the phase noise will then be $\delta\phi^{\left(1\right)}\rightarrow\epsilon^{\left(1\right)}10^{5}\sim10^{-4}$ for $s=0.01$ m which is sufficient to allow detections in this frequency spectrum. There is still the issue of the second order phase variations ($\delta\phi^{\left(2\right)}$). These can similarly be approximated, this time by two symmetric interferometers, now spread in the `$x$' direction, and taking the difference between them divided by the distance between them. As the two interferometers would have to be spread further apart to make room for the original interferometer then before they will only accurately measure linear change in $g$ across the interferometer. This suggest the error in the phase due to GGN after both methods of cancellations are used will be effectively the third order GGN effect, which will be a further $\frac{\Delta x}{\lambda}$ smaller than the second order effect, giving $\delta\phi^{\left(3\right)}\sim10^{-1}$ at $1$ mHz frequency. This is still significant and as such gravity gradient noise will create an effective noise floor to the sensitivity of our detector. To determine how it effects our sensitivity at other frequencies we use the scaling provided in~\cite{saulson1984terrestrial} of $1/\sqrt{f}$ to generate the noise floor after cancelling $\delta\phi^{\left(1\right)}$ and $\delta\phi^{\left(2\right)}$ as discussed above, for all relevant frequencies. The resulting GGN signal is then
\begin{eqnarray}
	\delta\phi^{\left(3\right)}&\sim&\left(\frac{\Delta x}{\lambda}\right)^2 \frac{2m}{\hbar}\left(0.6\frac{\sqrt{r_0}}{\sqrt{d}}\times\frac{3\times 10^{-11}}{\sqrt{f/1\textrm{ mHz}}}\right)\Delta x\tau_1 \\
	&\approx&\frac{8\times10^{-3}}{\sqrt{f/1\textrm{ Hz}}}
\end{eqnarray}
for the $m=10^{-17}$ kg interferometer used in figure~\ref{fig:GWStrainComparison}, this also shows that the optimal sensitivity here is in the $0.04$ Hz-$3$ Hz range. Note this also matches closely with the median GGN spectra given in~\cite{harms2015terrestrial}.

This is a somewhat crude model, treating both ground and atmospheric fluctuations at once, assuming uncorrelated fluctuations and integrating over each cell rather then summing. However as we are using actual measured results for $a_{\textrm{surface}}$ and in effect only concerned with the scaling with $r$ and $d$, we are unlikely to be lead astray by our model. Also we are using the measured median GGN spectra and as such likely over estimating the noise as it would actually effect our interferometer as we would intend for it to be placed at a `quiet' site with low GGN. Furthermore we can note that such a method of measuring and cancelling noise can be applied to other GW sensors, potentially extending the ground based observable frequencies in all GW sensors.

\subsection{Heisenberg Uncertainty Noise \label{subsec:Heisenberg uncertainty}}
Another key noise source in standard GW detectors is the fundamental noise due to the Heisenberg uncertainty limit. For simplicity we will consider the mass to be in a coherent state saturating the uncertainty principle, that is,
\begin{eqnarray}
\sigma_{x}\sigma_{p}&=&\frac{\hbar}{2}\\
\sigma_{p}&=&\sqrt{\frac{m\omega\hbar}{2}}\approx7\times10^{-24}\textrm{ kgms}^{-1}\\
\sigma_{x}&=&\sqrt{\frac{\hbar}{2m\omega}}\approx7\times10^{-12}\textrm{ m}.
\end{eqnarray}
where the particle is assumed to be released from a $100$ kHz trap. Beginning with position uncertainty there are two potential manners in which this could impact the final result, the first is uncertainty in the initial position giving
\begin{eqnarray}
\delta\left(\Delta\phi\left(h_{00}\right)\right)&\approx&\left(\frac{2mMG}{\hbar}a\tau_1^3\right)\left(\frac{1}{\left(R\right)^2}-\frac{1}{ \left(R+\sigma_{x}\right)^2}\right)\sin\left(\alpha\right)\\
&\approx&\left(\frac{4mMG\sigma_{x}}{\hbar R^{4}}a\tau_1^3\right)\sin\left(\alpha\right)
\sim10^{-7}\sin\left(\alpha\right)
\end{eqnarray}
where again $\alpha$ is the angle between the interferometer's $x$ axis and the local plane of constant Newtonian potential. The second manner in which position uncertainty due to HUP could manifest as noise is by impacting the overlap between the particle. However it is known~\cite{wan2016free} that this cannot effect the result as the phase difference is independent of the initial spacial state.

Along similar lines we can consider how the initial momentum HUP uncertainty results in phase uncertainty. This gives
\begin{equation}
\delta\left(\Delta\phi_{h_{00}}\right)\approx\frac{8MG\sigma_{p}a\tau_1^4}{\hbar R^3}\sin\left(\alpha\right) \sim10^{5}\sin\left(\alpha\right)
\end{equation}
as such provided $\alpha\ll10^{-5}$ this is also not an issue. As it is anticipated $\alpha\sim10^{-12}$~\cite{hogan2011precision} the HUP is not anticipated to be a limiting factor. 

\subsection{Particle-Particle Interactions \label{subsec:Particle-particle interactions}}

Any electrostatic interactions can be eliminated as the particle charge can be measured and modified down to the single electron level~\cite{frimmer2017}. The more concerning interactions will be particle-particle and particle-magnet interactions. The particle-particle interactions are kept in check by ensuring the particle flux is low  where here the flux is defined as the number of particles through the interferometer per second. The phase uncertainty it introduces is primarily due to the inter-particle Casimir interaction. It however can be minimised by ensuring a large enough $v_y$, for example, considering the effective Casimir potential ($U_{\textrm{C}}$) between two diamond ($\epsilon=5.7$) spheres of radius $\bar{R}$ a distance $d$ apart as
\begin{equation}
	U_{\textrm{C}}=\frac{23\hbar c \bar{R}^6}{4\pi d}\left(\frac{\epsilon-1}{\epsilon+2}\right)^2
\end{equation}
then provided $v_y=10$ ms$^{-1}$ then a flux $N=1000$ will lead to a phase uncertainty of approximately $0.002$ rad with a phase sensitivity to the $0.03$ radian level. When $v_y=1$ ms$^{-1}$ the highest allowable flux is about $N=90$ which gives a phase uncertainty of approximately $0.05$ rad with sensitivity of approximately $0.1$ rad. To this end we have considered $N=400$ with $v_y=10$ ms$^{-1}$ as sufficient to ensure the particle-particle interactions are negligible while also gaining phase sensitivity, with larger fluxes yielding phase sensitivity which would likely lost to other noises discussed above. Note that such large values for $v_y$ can be achieved for a polarizable particle (e.g. nanodiamond) using rapid acceleration in a pulsed optical field \cite{Barker2012}.

\subsection{Magnetic Field Fluctuations \label{subsec:Magnetic field fluctuations}}
Fluctuations of the magnetic field and its gradient will effect the interferometer in a number of differing ways: modify $\Delta x$, stop the interferometer closing perfectly and through the phase fluctuation associated with variations in the magnetic potential energy.

The source of the magnetic field fluctuations will be due to variations in the current through the wire taking $I\rightarrow I+\delta I$. Such fluctuations will translate to variations in the applied acceleration $\delta a$ given by
\begin{equation}
\frac{\delta a}{a}=\frac{\delta I}{I}
\end{equation}

Now if such fluctuations occur at time spans similar to the total interferometry time ($\tau_3=4\tau_1$) than they will automatically be cancelled to the alternating direction of the acceleration. Similarly if they occur much faster than again they will on average cancel throughout the interferometry process. As such the most significant position fluctuations occur if the sign of $\delta I$ changes at times $t=\tau_1$ and then again at $t=\tau_2$ suggesting a characteristic time span of $2\tau_1$ such that its contribution to the acceleration never cancels. In this instance we have 
\begin{equation}
\frac{\delta(\Delta x)}{\Delta x}=\frac{\delta a}{a}=\frac{\delta I}{I}
\end{equation}
although, if multiple particles are traversing the interferometer in series then for a later particle this effect would smaller or cancel completely, a detail we will neglect here so that we are not underestimating the noise. In the context of the Newtonian potential variations from one run to the next we found that we require $\delta(\Delta x)\le10$ nm (CF. Section~\ref{subsec:Gravitational signals as noise}), which, given the maximum superposition size is $\Delta x=1$ m, sets a limit to the current variation $\delta I = 10^{-8} I$ over a time-scale of $2\tau_1\approx1.5$ s. Now the ensure this is not exceeded the experimentalists would simply have to monitor the current with an ammeter to ensure drift is kept below this level.

Furthermore we can consider the current fluctuation due to thermal effects within the conducting wire by considering Johnson-Niquist noise which gives the current noise through the wire as

\begin{equation}
	\delta I=\sqrt{\frac{4 k_B T \Delta f}{\underline{R}}}
\end{equation}

where $k_B$ is Boltzmann's constant, $T$ the temperature of the wire, $\Delta f\sim1$ Hz the bandwidth for noise and $\underline{R}\sim 22$ k$\Omega$ is the resistance of the wire\cite{sugime2013CNT}. This gives a current noise of $\delta I\sim 10^{-12}$ A if the wire is maintained at room temperature. This is likely to be well below the required noise floor, even with the wire heating up well above room temperature.

This will also then lead to the particles overlapping only up to the bound given approximately by $\delta\left(\Delta x\right)$. However using the results derived below we can conclude $\delta\left(\Delta x\right)\sim10^{-15}$ m, far below the assumed wavepacket spread due to Heisenberg uncertainty of $\sigma_{x}\sim10^{-11}$m and so is not of significant concern.

Finally we can consider the phase fluctuation due to the magnetic field coupling. This phase due to the coupling between an electronic spin and an aligned magnetic field is given by

\begin{eqnarray}
	\phi_{\vec{B}}&=&\frac{\vec{\mu}\cdot\vec{B}t}{\hbar} \nonumber\\
	&=&-\frac{e g S}{2m_e \hbar}\frac{\mu_0I}{2\pi D}t
\end{eqnarray}
where $e$ and $m_e$ are the charge and mass of an electron, $g\approx2$ is the gyromagnetic ratio and $S$ is the spin angular momentum. Now as the spin state is reversed throughout the interferometer, the total phase will effectively unwind itself, up to the stability in both the mean magnetic field strength and timing accuracy. In this way the phase difference will be $\Delta\phi_{\vec{B}}=0$ up to some stochastic fluctuations given by

\begin{eqnarray}
	\delta\left(\Delta\phi_{\vec{B}}\right)&=&\frac{e g \hbar}{2m_e \hbar}\times\frac{\mu_0\delta I}{2\pi D}\tau_3+\frac{e g \hbar}{2m_e \hbar}\times\frac{\mu_0\rho_I\pi D^2}{2\pi D}\delta t\nonumber\\
	&\sim&\frac{10^{-7}\textrm{m}}{D} + 10^{17}\textrm{m}^{-1}\textrm{s}^{-1} D\delta t
\end{eqnarray}

Now the first term implies a restriction on the distance between the centre of the wire and the particle of $D\ge1~\mu$m while the second term implies a limit on the timing uncertainty of $\delta t\ll 10^{17} D$ m$^{-1}$s. So, taking $D=2\times10^{-5}$ m, thus requiring a current of $I\approx2000$ A and magnetic field magnitude of $B=40$ T, a timing uncertainty of $\delta t\le10^{-13}$ s is required. This is certainly a difficult requirement, but does not seem completely unreasonable given the historical achievement of pico-second ($10^{-12}$ s) timings with microwave lasers~\cite{mourou1981picosecond} with femto-second also achieved more recently~\cite{kim2008drift}. 

Each small section of current carrying wire pair will have to be controlled independently and thus will have an independent current fluctuating stochastically about the intended value $I$. Therefore, there is no independent noise at frequencies lower than that which corresponds to the time each wire pair controls the particle -- noise at such frequencies essentially corresponds to the sum of noises from blocks of consecutive wires. Thus we do not need to consider them separately; considering the noise at the frequency corresponding to the time each wire pair controls the particle suffices. In this case, the wire pair controls the motion of the particle for typically $t_{\textrm{wire}}=7$ $\mu$s to ensure the particle sees a uniform, linear magnetic field gradient throughout the interferometry process. This corresponds to a noise frequency of $f_{\textrm{wire}} \sim 1.4\times 10^5$ Hz. Over the total time of the experiment $\sim 1$s, the uncertain part of the Zeeman phase accumulated will be a summative random walk type phase. Here each wire interval is responsible for a step in the random walk. For this to be negligible, we require the random part of the magnetic field magnitude at the frequency $f_{\textrm{wire}}$ to be $\delta B(f_{\textrm{wire}}) < \frac{\hbar}{\mu_B} \sqrt{f_{\textrm{wire}}}\sim 4$ nT 
(alternatively, simply keeping track of the magnetic field fluctuations to this accuracy will suffice). This corresponds to a current uncertainty of $\delta I(f_{\textrm{wire}}) < 20 \mu A$ at the frequency of $f_{\textrm{wire}}$. For frequencies $f > f_{\textrm{wire}}$, the constraint on $\delta B(f) \leq 4\sqrt{f/f_{\textrm{wire}}} $ nT will only be easier to satisfy. Additionally, the fluctuation in the gradient will also cause an uncertainty in the particle's position of
\begin{eqnarray}
\delta\left(\Delta x\right)&=&\frac{\delta I t_{\textrm{wire}}^2 \Delta x}{I \tau_1^2} \nonumber\\
&\approx &\frac{\delta I \times 10^{-10}}{I}
\end{eqnarray}
which, by requiring $\delta\left(\Delta x\right)\ll\sigma_{x}$, bounds the high frequency ($MHz$) magnetic field fluctuations $\delta I$ to $\delta I\ll20$ A. This can be extrapolated to give a general bound, frequency dependent bound of
\begin{equation}
\delta I(f)\le\frac{2\times 10^{3}\textrm{A Hz}^{-1/2}}{\sqrt{f}}
\end{equation}

\subsection{Particle-Magnet Casimir Interaction \label{subsec:Particle-Magnet interactions}}

To model the particle-magnet Casimir induced phase fluctuations we can note that, as the particle radius is $\bar{R}\sim10^{-7}$ m and the particle-magnet surface distance is kept at $\Lambda=10^{-5}$ m, the particle-magnet system can be considered to be in the long range limit, the path phase difference of~\cite{Ford1998Casimir}
\begin{equation}
\Delta \phi_{\textrm{Casimir}}=\frac{23 c \bar{R}^3}{4\pi \Lambda^4}\tau_3 \sim 10^6\textrm{~rad}
\end{equation}
where $c$ is the speed of light and $\tau_3$ is the total interferometry time as shown in Figure~\ref{fig:Interferometer path diagram}. While this is significant, it is a constant phase provided the separation distance is also kept constant it can be normalised for in the output. This however requires certainty in the particle-magnet separation to be $\sim10^{-11}$ m while the aforementioned timing stability is sufficient here.  This also leads to a maximum path displacement of $\sim10^{-3}$ m over the length of the interferometer leading to the two state not overlapping without also adjusting the spin-$0$ arm of the interferometer. This displacement will be stable to the same level as the phase however and so should not limit the ability to completely overlap the two.

Patch potentials refer to electrostatic interactions between regions of non-zero charge on a globally charge neutral object. The patch potential interactions can largely be dealt with by using single crystal particles as the interferometry masses. The particle-magnet patch potential interaction will be further minimised by the geometry of the system. The patch potential force~\cite{trenkel2003PatchPotentials} scales as 
\begin{equation}
F\propto \frac{\bar{R}e^{\Lambda/a}}{\sinh(\Lambda/a)} 
\end{equation}
where again $\bar{R}\sim10^{-7}$ m is the particle radius, $\Lambda\sim10^{-4}$ m is the particle-magnet separation and here $a<\bar{R}$ is the size of the patch potential. This exponential suppression means that the patch potential is effectively negligible. Furthermore since the particle can me moves along the magnet, and by initialising the particles as physically spinning any patch potential interactions can be further averaged out. Finally if the particle is constructed out of a single crystal these patch potential effects would negligible.

\section{Conclusion \label{sec:Conclusion}}
We have presented the possibility of using the interferometry of mesoscopic objects (say, objects of mass $\sim 10^{-17}$ kg) to detect both first and second order derivatives of the space-time metric in a compact setup. We have found that for mesoscopic masses, such interferometry is not only sensitive to the Newtonian potential, Earth's frame dragging, but also extremely weak signals such as mid frequency GWs for a ground based detector, and low frequency GWs for a space based detector. We have presented an example form for such a mesoscopic mass interferometer and presented the expected sensitivity for our device. In designing our example detector, we have identified the requirements which must be met to mitigate the known sources of noise, such as GGN, uncertainty principle, Casimir and patch-potential interactions. The SG principle of the specific interferometer design implies that simply by changing the orientation of a magnet, the whole interferometer is re-oriented to both identify the angular origin of sources and couple to different components of the metric tensor. Furthermore, the manner in which the phase difference accumulated due to the Newtonian potential and GW signals scale with the experimental parameters $a$ and $\tau_1$ points to an important and fundamental difference between this type of interferometer and light based interferometers. This difference provides an avenue to further improve GW sensitivity while reducing many noise sources, including GGN. The compactness means that whole GW sensitive interferometers can be put in a single vibrational isolation platform \cite{Martynov:2016fzi} and large networks of interferometers can be built to identify and cancel noise. Less demanding values for $\partial_{x}\vec{B}$ and the coherence times suffice to detect the less demanding components such as $h_{00}$ or for accelerometry (e.g. $\partial_{x}\vec{B}=10^4$ Tm$^{-1}$, $\tau_1\sim 70$ ms, $10^{-18}$ kgs and $\Delta x = 1$ mm can already detect both the Newtonian curvature and the Earth's frame dragging). Attempts to build the most ambitious limit of the interferometers, namely, for GW detection with superpositions of $10^{-17}$ kg masses as discussed here will also push the limits of macroscopicity of superpositions as defined in~\cite{nimmrichter2013macroscopicity} to $\mu\approx26$ (where atomic and macromolecular interferometry have achieved $\mu\approx11$ \cite{kovachy2015quantum} and $\mu\approx 14.5$ \cite{nimmrichter2011testing} respectively, with an actual Schroedinger cat experiment corresponding to $\mu\approx57$). This will constrain intrinsic collapse models \cite{penrose1996gravity, kovachy2015quantum} to an electron coherence time  $\tau_e\sim\times10^{26}$ s at a critical length scale $\hbar/\sigma_q\sim 1$ m. We may be able to test short distance modifications of gravity ~\cite{Biswas:2011ar,Biswas:2005qr}, and the gravitational self-localization of wavefunctions ~\cite{Buoninfante:2017kgj,Buoninfante:2017rbw}.

\ack AM's research is funded by the Netherlands Organisation for Scientific Research (NWO) grant number 680-91-119, GWM is supported by the Royal Society. We acknowledge EPSRC grant No. EP/M013243/1. PFB and SB would like to acknowledge EPSRC grants No. EP/N031105/1 and EP/S000267/1. We would also like to thank Ron Folman for his insightful discussions regarding the experimental implementation of large magnetic field gradients using current carrying wires.

\section*{References}
\bibliography{bibliography}

\begin{thebibliography}{100}

\bibitem{RevModPhys.81.1051}
Alexander~D. Cronin, J\"org Schmiedmayer, and David~E. Pritchard.
\newblock Optics and interferometry with atoms and molecules.
\newblock {\em Rev. Mod. Phys.}, 81:1051--1129, Jul 2009.

\bibitem{gerlich2007kapitza}
Stefan Gerlich, Lucia Hackerm{\"u}ller, Klaus Hornberger, Alexander Stibor,
  Hendrik Ulbricht, Michael Gring, Fabienne Goldfarb, Tim Savas, Marcel
  M{\"u}ri, Marcel Mayor, et~al.
\newblock A kapitza--dirac--talbot--lau interferometer for highly polarizable
  molecules.
\newblock {\em Nature Physics}, 3(10):711, 2007.

\bibitem{bose1999scheme}
Sougato Bose, Kurt Jacobs, and Peter~L Knight.
\newblock Scheme to probe the decoherence of a macroscopic object.
\newblock {\em Physical Review A}, 59(5):3204, 1999.

\bibitem{armour2002entanglement}
A.~D. Armour, M.~P. Blencowe, and Keith~C Schwab.
\newblock Entanglement and decoherence of a micromechanical resonator via
  coupling to a cooper-pair box.
\newblock {\em Physical Review Letters}, 88(14):148301, 2002.

\bibitem{marshall2003towards}
William Marshall, Christoph Simon, Roger Penrose, and Dik Bouwmeester.
\newblock Towards quantum superpositions of a mirror.
\newblock {\em Physical Review Letters}, 91(13):130401, 2003.

\bibitem{sekatski2014macroscopic}
Pavel Sekatski, Markus Aspelmeyer, and Nicolas Sangouard.
\newblock Macroscopic optomechanics from displaced single-photon entanglement.
\newblock {\em Physical Review Letters}, 112(8):080502, 2014.

\bibitem{romero2010toward}
Oriol Romero-Isart, Mathieu~L Juan, Romain Quidant, and J~Ignacio Cirac.
\newblock Toward quantum superposition of living organisms.
\newblock {\em New Journal of Physics}, 12(3):033015, 2010.

\bibitem{romero2011large}
Oriol Romero-Isart, Anika~C Pflanzer, Florian Blaser, Rainer Kaltenbaek,
  Nikolai Kiesel, Markus Aspelmeyer, and J~Ignacio Cirac.
\newblock Large quantum superpositions and interference of massive
  nanometer-sized objects.
\newblock {\em Physical Review Letters}, 107(2):020405, 2011.

\bibitem{khalili2010preparing}
Farid Khalili, Stefan Danilishin, Haixing Miao, Helge M{\"u}ller-Ebhardt, Huan
  Yang, and Yanbei Chen.
\newblock Preparing a mechanical oscillator in non-gaussian quantum states.
\newblock {\em Physical Review Letters}, 105(7):070403, 2010.

\bibitem{scala2013matter}
Matteo Scala, M.~S. Kim, G.~W. Morley, P.~F. Barker, and S.~Bose.
\newblock Matter-wave interferometry of a levitated thermal nano-oscillator
  induced and probed by a spin.
\newblock {\em Physical Review Letters}, 111(18):180403, 2013.

\bibitem{bateman2014near}
James Bateman, Stefan Nimmrichter, Klaus Hornberger, and Hendrik Ulbricht.
\newblock Near-field interferometry of a free-falling nanoparticle from a
  point-like source.
\newblock {\em Nature Communications}, 5:4788, 2014.

\bibitem{yin2013large}
Zhang~Q. Yin, Tongcang Li, Xiang Zhang, and L.~M. Duan.
\newblock Large quantum superpositions of a levitated nanodiamond through
  spin-optomechanical coupling.
\newblock {\em Physical Review A}, 88(3):033614, 2013.

\bibitem{pino2018chip}
H~Pino, J~Prat-Camps, K~Sinha, B~Prasanna Venkatesh, and O~Romero-Isart.
\newblock On-chip quantum interference of a superconducting microsphere.
\newblock {\em Quantum Science and Technology}, 3(2):025001, 2018.

\bibitem{clarke2018growingPublished}
Jack Clarke and Michael~R Vanner.
\newblock Growing macroscopic superposition states via cavity quantum
  optomechanics.
\newblock {\em Quantum Science and Technology}, 4(1):014003, 2019.

\bibitem{ringbauer2016generation}
M~Ringbauer, TJ~Weinhold, AG~White, and MR~Vanner.
\newblock Generation of mechanical interference fringes by multi-photon quantum
  measurement.
\newblock {\em arXiv preprint arXiv:1602.05955}, 2016.

\bibitem{khosla2018displacemon}
Kiran~E Khosla, Michael~R Vanner, Natalia Ares, and Edward~Alexander Laird.
\newblock Displacemon electromechanics: how to detect quantum interference in a
  nanomechanical resonator.
\newblock {\em Physical Review X}, 8(2):021052, 2018.

\bibitem{kaltenbaek2012MAQRO}
Rainer Kaltenbaek, Gerald Hechenblaikner, Nikolai Kiesel, Oriol Romero-Isart,
  Keith~C Schwab, Ulrich Johann, and Markus Aspelmeyer.
\newblock Macroscopic quantum resonators (maqro).
\newblock {\em Experimental Astronomy}, 34(2):123--164, 2012.

\bibitem{wan2016free}
C.~Wan, M.~Scala, G.~W. Morley, {ATM}~A Rahman, H~Ulbricht, J~Bateman, P.~F.
  Barker, S.~Bose, and M.~S. Kim.
\newblock Free nano-object ramsey interferometry for large quantum
  superpositions.
\newblock {\em Physical Review Letters}, 117(14):143003, 2016.

\bibitem{romero2017coherent}
Oriol Romero-Isart.
\newblock Coherent inflation for large quantum superpositions of levitated
  microspheres.
\newblock {\em New Journal of Physics}, 19(12):123029, 2017.

\bibitem{penrose1996gravity}
Roger Penrose.
\newblock On gravity's role in quantum state reduction.
\newblock {\em General relativity and gravitation}, 28(5):581--600, 1996.

\bibitem{bassi2013models}
Angelo Bassi, Kinjalk Lochan, Seema Satin, Tejinder~P Singh, and Hendrik
  Ulbricht.
\newblock Models of wave-function collapse, underlying theories, and
  experimental tests.
\newblock {\em Reviews of Modern Physics}, 85(2):471, 2013.

\bibitem{Bose:2017nin}
Sougato Bose, Anupam Mazumdar, Gavin~W. Morley, Hendrik Ulbricht, Marko
  Toro\u{s}, Mauro Paternostro, Andrew~A. Geraci, Peter~F. Barker, M.~S. Kim,
  and Gerard Milburn.
\newblock {Spin Entanglement Witness for Quantum Gravity}.
\newblock {\em Phys. Rev. Lett.}, 119(24):240401, 2017.

\bibitem{Marletto2017}
C.~Marletto and V.~Vedral.
\newblock Gravitationally induced entanglement between two massive particles is
  sufficient evidence of quantum effects in gravity.
\newblock {\em Phys. Rev. Lett.}, 119:240402, Dec 2017.

\bibitem{Colella1975}
Roberto Colella, Albert~W Overhauser, and Samuel~A Werner.
\newblock Observation of gravitationally induced quantum interference.
\newblock {\em Physical Review Letters}, 34(23):1472, 1975.

\bibitem{Anandan1984}
Jeeva Anandan.
\newblock Effect of newtonian gravitational potential on a superfluid josephson
  interferometer.
\newblock {\em Physical Review B}, 30(7):3717, 1984.

\bibitem{Mcguirk2002}
Jeffrey~Michael McGuirk, G.~T. Foster, J.~B. Fixler, M.~J. Snadden, and M.~A.
  Kasevich.
\newblock Sensitive absolute-gravity gradiometry using atom interferometry.
\newblock {\em Physical Review A}, 65(3):033608, 2002.

\bibitem{qvarfort2018gravimetryPublished}
Sofia Qvarfort, Alessio Serafini, Peter~F Barker, and Sougato Bose.
\newblock Gravimetry through non-linear optomechanics.
\newblock {\em Nature Communications}, 9(1):3690, 2018.

\bibitem{Armata2017}
Federico Armata, Ludovico Latmiral, A.~D.~K Plato, and M.~S. Kim.
\newblock Quantum limits to gravity estimation with optomechanics.
\newblock {\em Physical Review A}, 96(4):043824, 2017.

\bibitem{asenbaum2017phase}
Peter Asenbaum, Chris Overstreet, Tim Kovachy, Daniel~D Brown, Jason~M Hogan,
  and Mark~A Kasevich.
\newblock Phase shift in an atom interferometer due to spacetime curvature
  across its wave function.
\newblock {\em Physical review letters}, 118(18):183602, 2017.

\bibitem{WSC(1979)}
S.~A. Werner, J.~L. Staudenmann, and R.~Colella.
\newblock Effect of earth's rotation on the quantum mechanical phase of the
  neutron.
\newblock {\em Phys. Rev. Lett.}, 42:1103--1106, Apr 1979.

\bibitem{Anandan1977}
J~Anandan.
\newblock Gravitational and rotational effects in quantum interference.
\newblock {\em Physical Review D}, 15(6):1448, 1977.

\bibitem{Dimopoulos2008a}
Savas Dimopoulos, Peter~W Graham, Jason~M Hogan, and Mark~A Kasevich.
\newblock General relativistic effects in atom interferometry.
\newblock {\em Physical Review D}, 78(4):042003, 2008.

\bibitem{GravityProbeB}
C.~W.~F. Everitt, D.~B. DeBra, B.~W. Parkinson, J.~P. Turneaure, J.~W. Conklin,
  M.~I. Heifetz, G.~M. Keiser, A.~S. Silbergleit, T.~Holmes, J.~Kolodziejczak,
  M.~Al-Meshari, J.~C. Mester, B.~Muhlfelder, V.~G. Solomonik, K.~Stahl, P.~W.
  Worden, W.~Bencze, S.~Buchman, B.~Clarke, A.~Al-Jadaan, H.~Al-Jibreen, J.~Li,
  J.~A. Lipa, J.~M. Lockhart, B.~Al-Suwaidan, M.~Taber, and S.~Wang.
\newblock Gravity probe b: Final results of a space experiment to test general
  relativity.
\newblock {\em Phys. Rev. Lett.}, 106:221101, May 2011.

\bibitem{roura2017circumventing}
Albert Roura.
\newblock Circumventing heisenberg’s uncertainty principle in atom
  interferometry tests of the equivalence principle.
\newblock {\em Physical review letters}, 118(16):160401, 2017.

\bibitem{roura2018gravitational}
Albert Roura.
\newblock Gravitational redshift in quantum-clock interferometry.
\newblock {\em arXiv:1810.06744}, 2018.

\bibitem{Abbott2016}
Benjamin~P Abbott, Richard Abbott, TD~Abbott, MR~Abernathy, Fausto Acernese,
  Kendall Ackley, Carl Adams, Thomas Adams, Paolo Addesso, RX~Adhikari, et~al.
\newblock Observation of gravitational waves from a binary black hole merger.
\newblock {\em Physical Review Letters}, 116(6):061102, 2016.

\bibitem{Harry2010}
Gregory~M Harry, LIGO~Scientific Collaboration, et~al.
\newblock Advanced ligo: the next generation of gravitational wave detectors.
\newblock {\em Classical and Quantum Gravity}, 27(8):084006, 2010.

\bibitem{LISA}
Pau Amaro-Seoane, Heather Audley, Stanislav Babak, John Baker, Enrico Barausse,
  Peter Bender, Emanuele Berti, Pierre Binetruy, Michael Born, Daniele
  Bortoluzzi, et~al.
\newblock Laser interferometer space antenna.
\newblock {\em arXiv preprint arXiv:1702.00786}, 2017.

\bibitem{Chiao2004}
Raymond~Y Chiao and Achilles~D Speliotopoulos.
\newblock Towards migo, the matter-wave interferometric gravitational-wave
  observatory, and the intersection of quantum mechanics with general
  relativity.
\newblock {\em Journal of Modern Optics}, 51(6-7):861--899, 2004.

\bibitem{Roura2006}
Albert Roura, Dieter~R. Brill, B.~L. Hu, Charles~W. Misner, and William~D.
  Phillips.
\newblock Gravitational wave detectors based on matter wave interferometers
  (migo) are no better than laser interferometers (ligo).
\newblock {\em Phys. Rev. D}, 73:084018, Apr 2006.

\bibitem{Foffa2006}
Stefano Foffa, Alice Gasparini, Michele Papucci, and Riccardo Sturani.
\newblock Sensitivity of a small matter-wave interferometer to gravitational
  waves.
\newblock {\em Physical Review D}, 73(2):022001, 2006.

\bibitem{Dimopoulos2008}
Savas Dimopoulos, Peter~W. Graham, Jason~M. Hogan, Mark~A. Kasevich, and
  Surjeet Rajendran.
\newblock An atomic gravitational wave interferometric sensor (agis).
\newblock {\em Physical Review D}, 78(12):122002, 2008.

\bibitem{Dimopoulos2009}
Savas Dimopoulos, Peter~W Graham, Jason~M Hogan, Mark~A Kasevich, and Surjeet
  Rajendran.
\newblock Gravitational wave detection with atom interferometry.
\newblock {\em Physics Letters B}, 678(1):37--40, 2009.

\bibitem{MIGA}
B.~Canuel, A.~Bertoldi, L.~Amand, E.~Pozzo~di Borgo, T.~Chantrait,
  C.~Danquigny, M.~Dovale~{\'A}lvarez, B.~Fang, A.~Freise, R.~Geiger,
  J.~Gillot, S.~Henry, J.~Hinderer, D.~Holleville, J.~Junca, G.~Lef{\`e}vre,
  M.~Merzougui, N.~Mielec, T.~Monfret, S.~Pelisson, M.~Prevedelli, S.~Reynaud,
  I.~Riou, Y.~Rogister, S.~Rosat, E.~Cormier, A.~Landragin, W.~Chaibi,
  S.~Gaffet, and P.~Bouyer.
\newblock Exploring gravity with the miga large scale atom interferometer.
\newblock {\em Scientific Reports}, 8(1):14064, 2018.

\bibitem{MAGIS}
Peter~W Graham, Jason~M Hogan, Mark~A Kasevich, Surjeet Rajendran, and Roger~W
  Romani.
\newblock Mid-band gravitational wave detection with precision atomic sensors.
\newblock {\em arXiv preprint arXiv:1711.02225}, 2017.

\bibitem{Raetzel2017}
Dennis R\"{a}tzel, Fabienne Schneiter, Daniel Braun, Tupac Bravo, Richard Howl,
  Maximilian P.~E. Lock, and Ivette Fuentes.
\newblock Frequency spectrum of an optical resonator in a curved spacetime.
\newblock {\em New Journal of Physics}, 20(5):053046, 2018.

\bibitem{Raetzel2018}
Dennis R\"{a}tzel, Richard Howl, Joel Lindkvist, and Ivette Fuentes.
\newblock Dynamical response of bose–einstein condensates to oscillating
  gravitational fields.
\newblock {\em New Journal of Physics}, 20(7):073044, 2018.

\bibitem{Geraci2013}
Asimina Arvanitaki and Andrew~A. Geraci.
\newblock Detecting high-frequency gravitational waves with optically levitated
  sensors.
\newblock {\em Phys. Rev. Lett.}, 110:071105, Feb 2013.

\bibitem{pontin2018levitated}
Antonio Pontin, Lauren~S Mourounas, Andrew~A Geraci, and Peter~F Barker.
\newblock Levitated optomechanics with a fiber fabry--perot interferometer.
\newblock {\em New Journal of Physics}, 20(2):023017, 2018.

\bibitem{ando2010torsion}
Masaki Ando, Koji Ishidoshiro, Kazuhiro Yamamoto, Kent Yagi, Wataru Kokuyama,
  Kimio Tsubono, and Akiteru Takamori.
\newblock Torsion-bar antenna for low-frequency gravitational-wave
  observations.
\newblock {\em Physical review letters}, 105(16):161101, 2010.

\bibitem{Folman2013}
Shimon Machluf, Yonathan Japha, and Ron Folman.
\newblock Coherent stern--gerlach momentum splitting on an atom chip.
\newblock {\em Nature Communications}, 4:2424, 09 2013.

\bibitem{Folman2018}
Yair Margalit, Zhifan Zhou, Or~Dobkowski, Yonathan Japha, Daniel Rohrlich,
  Samuel Moukouri, and Ron Folman.
\newblock Realization of a complete stern-gerlach interferometer.
\newblock {\em arXiv preprint arXiv:1801.02708}, 2018.

\bibitem{folman2019}
O.~Amit, Y.~Margalit, O.~Dobkowski, Z.~Zhou, Y.~Japha, M.~Zimmermann, M.~A.
  Efremov, F.~A. Narducci, E.~M. Rasel, W.~P. Schleich, and R.~Folman.
\newblock ${T}^{3}$ stern-gerlach matter-wave interferometer.
\newblock {\em Phys. Rev. Lett.}, 123:083601, Aug 2019.

\bibitem{hammerer2015optomechanics}
Klemens Hammerer and Markus Aspelmeyer.
\newblock Optomechanics: Diamonds take off.
\newblock {\em Nature Photonics}, 9(10):633, 2015.

\bibitem{neukirch2015multi}
Levi~P Neukirch, Eva Von~Haartman, Jessica~M Rosenholm, and A~Nick Vamivakas.
\newblock Multi-dimensional single-spin nano-optomechanics with a levitated
  nanodiamond.
\newblock {\em Nature Photonics}, 9(10):653, 2015.

\bibitem{siyushev2014coherent}
P~Siyushev, K~Xia, R~Reuter, M~Jamali, N~Zhao, N~Yang, C~Duan, N~Kukharchyk,
  AD~Wieck, R~Kolesov, et~al.
\newblock Coherent properties of single rare-earth spin qubits.
\newblock {\em Nature communications}, 5:3895, 2014.

\bibitem{liu2018controlled}
Shuping Liu, Diana Serrano, Alexandre Fossati, Alexandre Tallaire, Alban
  Ferrier, and Philippe Goldner.
\newblock Controlled size reduction of rare earth doped nanoparticles for
  optical quantum technologies.
\newblock {\em RSC advances}, 8(65):37098--37104, 2018.

\bibitem{frimmer2017}
Martin Frimmer, Karol Luszcz, Sandra Ferreiro, Vijay Jain, Erik Hebestreit, and
  Lukas Novotny.
\newblock Controlling the net charge on a nanoparticle optically levitated in
  vacuum.
\newblock {\em Phys. Rev. A}, 95:061801(R), Jun 2017.

\bibitem{deli2019motional}
Uro{\v{s}} Deli{\'c}, Manuel Reisenbauer, Kahan Dare, David Grass, Vladan
  Vuleti{\'c}, Nikolai Kiesel, and Markus Aspelmeyer.
\newblock Motional quantum ground state of a levitated nanoparticle from room
  temperature, 2019.

\bibitem{chang2010cavity}
Darrick~E Chang, CA~Regal, SB~Papp, DJ~Wilson, J~Ye, O~Painter, H~Jeff Kimble,
  and P~Zoller.
\newblock Cavity opto-mechanics using an optically levitated nanosphere.
\newblock {\em Proceedings of the National Academy of Sciences},
  107(3):1005--1010, 2010.

\bibitem{ferialdi2019optimal}
Luca Ferialdi, Ashley Setter, Marko Toro{\v{s}}, Chris Timberlake, and Hendrik
  Ulbricht.
\newblock Optimal control for feedback cooling in cavityless levitated
  optomechanics.
\newblock {\em New Journal of Physics}, 2019.

\bibitem{rahman2017laser}
ATM~Anishur Rahman and PF~Barker.
\newblock Laser refrigeration, alignment and rotation of levitated yb 3+: Ylf
  nanocrystals.
\newblock {\em Nature Photonics}, 11(10):634, 2017.

\bibitem{mungan1999laser}
Carl~E Mungan and Timothy~R Gosnell.
\newblock Laser cooling of solids.
\newblock In {\em Advances in Atomic, Molecular, and Optical Physics},
  volume~40, pages 161--228. Elsevier, 1999.

\bibitem{henkel2019stern}
Carsten Henkel, Georg Jacob, Felix Stopp, Ferdinand Schmidt-Kaler, Mark Keil,
  Yonathan Japha, and Ron Folman.
\newblock Stern--gerlach splitting of low-energy ion beams.
\newblock {\em New Journal of Physics}, 21(8):083022, 2019.

\bibitem{Visser2018}
Matt Visser.
\newblock Post-newtonian particle physics in curved spacetime.
\newblock {\em arXiv preprint arXiv:1802.00651}, 2018.

\bibitem{Weinberg1972}
Steven Weinberg.
\newblock {\em Gravitation and cosmology: principles and applications of the
  general theory of relativity}, volume~1.
\newblock Wiley New York, 1972.

\bibitem{Martynov:2016fzi}
Benjamin~P. Abbott et~al.
\newblock {Sensitivity of the Advanced LIGO detectors at the beginning of
  gravitational wave astronomy}.
\newblock {\em Phys. Rev.}, D93(11):112004, 2016.
\newblock [Addendum: Phys. Rev.D97,no.5,059901(2018)].

\bibitem{ellis2018maximal}
John Ellis, Marek Lewicki, and Jos{\'e}~Miguel No.
\newblock On the maximal strength of a first-order electroweak phase transition
  and its gravitational wave signal.
\newblock {\em arXiv preprint arXiv:1809.08242}, 2018.

\bibitem{pedernales2019motional}
Julen~S Pedernales, Gavin~W Morley, and Martin~B Plenio.
\newblock Motional dynamical decoupling for matter-wave interferometry.
\newblock {\em arXiv preprint arXiv:1906.00835}, 2019.

\bibitem{subramaniam2013one}
Chandramouli Subramaniam, Takeo Yamada, Kazufumi Kobashi, Atsuko Sekiguchi,
  Don~N Futaba, Motoo Yumura, and Kenji Hata.
\newblock One hundred fold increase in current carrying capacity in a carbon
  nanotube--copper composite.
\newblock {\em Nature communications}, 4(1):1--7, 2013.

\bibitem{Gabrielse2016}
DW~Fitzakerley, MC~George, EA~Hessels, TDG Skinner, CH~Storry, M~Weel,
  G~Gabrielse, CD~Hamley, N~Jones, K~Marable, et~al.
\newblock Electron-cooled accumulation of 4$\times$ 109 positrons for
  production and storage of antihydrogen atoms.
\newblock {\em Journal of Physics B: Atomic, Molecular and Optical Physics},
  49(6):064001, 2016.

\bibitem{bose2018matter}
Sougato Bose and Gavin~W Morley.
\newblock Matter and spin superposition in vacuum experiment (massive).
\newblock {\em arXiv preprint arXiv:1810.07045}, 2018.

\bibitem{Romero-Isart2011}
Oriol Romero-Isart.
\newblock Quantum superposition of massive objects and collapse models.
\newblock {\em Physical Review A}, 84(5):052121, 2011.

\bibitem{Bar-Gill(2013)}
N.~Bar-Gill, L.~M. Pham, A.~Jarmola, D.~Budker, and R.~L. Walsworth.
\newblock Solid-state electronic spin coherence time approaching one second.
\newblock {\em Nature Communications}, 4:1743 EP --, 04 2013.

\bibitem{Abobeih(2018)}
M.~H. Abobeih, J.~Cramer, M.~A. Bakker, N.~Kalb, M.~Markham, D.~J. Twitchen,
  and T.~H. Taminiau.
\newblock One-second coherence for a single electron spin coupled to a
  multi-qubit nuclear-spin environment.
\newblock {\em Nature Communications}, 9(1):2552, June 2018.

\bibitem{knowles2014observing}
Helena~S Knowles, Dhiren~M Kara, and Mete Atat{\"u}re.
\newblock Observing bulk diamond spin coherence in high-purity nanodiamonds.
\newblock {\em Nature materials}, 13(1):21, 2014.

\bibitem{steane2017matter}
Andrew~M Steane.
\newblock Matter-wave coherence limit owing to cosmic gravitational wave
  background.
\newblock {\em Physics Letters A}, 381(47):3905--3908, 2017.

\bibitem{hughes1998seismic}
Scott~A Hughes and Kip~S Thorne.
\newblock Seismic gravity-gradient noise in interferometric gravitational-wave
  detectors.
\newblock {\em Physical Review D}, 58(12):122002, 1998.

\bibitem{moore2014gravitational}
Christopher~J. Moore, Robert~H. Cole, and Christopher P.~L. Berry.
\newblock Gravitational-wave sensitivity curves.
\newblock {\em Classical and Quantum Gravity}, 32(1):015014, 2014.

\bibitem{hogan2011precision}
JM~Hogan, J~Hammer, S-W Chiow, S~Dickerson, DMS Johnson, T~Kovachy,
  A~Sugarbaker, and MA~Kasevich.
\newblock Precision angle sensor using an optical lever inside a sagnac
  interferometer.
\newblock {\em Optics letters}, 36(9):1698--1700, 2011.

\bibitem{PhysRevD.88.122003}
Jan Harms, Bram J.~J. Slagmolen, Rana~X. Adhikari, M.~Coleman Miller, Matthew
  Evans, Yanbei Chen, Holger M\"uller, and Masaki Ando.
\newblock Low-frequency terrestrial gravitational-wave detectors.
\newblock {\em Phys. Rev. D}, 88:122003, Dec 2013.

\bibitem{Geiger2016}
W.~Chaibi, R.~Geiger, B.~Canuel, A.~Bertoldi, A.~Landragin, and P.~Bouyer.
\newblock Low frequency gravitational wave detection with ground-based atom
  interferometer arrays.
\newblock {\em Phys. Rev. D}, 93:021101(R), Jan 2016.

\bibitem{harms2009simulation}
Jan Harms, Riccardo DeSalvo, Steven Dorsher, and Vuk Mandic.
\newblock Simulation of underground gravity gradients from stochastic seismic
  fields.
\newblock {\em Physical Review D}, 80(12):122001, 2009.

\bibitem{saulson1984terrestrial}
Peter~R Saulson.
\newblock Terrestrial gravitational noise on a gravitational wave antenna.
\newblock {\em Physical Review D}, 30(4):732, 1984.

\bibitem{pitkin2011gravitational}
Matthew Pitkin, Stuart Reid, Sheila Rowan, and Jim Hough.
\newblock Gravitational wave detection by interferometry (ground and space).
\newblock {\em Living reviews in relativity}, 14(1):5, 2011.

\bibitem{harms2015terrestrial}
Jan Harms.
\newblock Terrestrial gravity fluctuations.
\newblock {\em Living reviews in relativity}, 18(1):3, 2015.

\bibitem{nishida2017ambient}
Kiwamu Nishida.
\newblock Ambient seismic wave field.
\newblock {\em Proceedings of the Japan Academy, Series B}, 93(7):423--448,
  2017.

\bibitem{Barker2012}
C.~Maher-McWilliams, P.~Douglas, and P.~F. Barker.
\newblock Laser-driven acceleration of neutral particles.
\newblock {\em Nature Photonics}, 6:386 EP --, 04 2012.

\bibitem{sugime2013CNT}
Hisashi Sugime, Santiago Esconjauregui, Junwei Yang, Lorenzo D'Arsi{\'e},
  Rachel~A Oliver, Sunil Bhardwaj, Cinzia Cepek, and John Robertson.
\newblock Low temperature growth of ultra-high mass density carbon nanotube
  forests on conductive supports.
\newblock {\em Applied Physics Letters}, 103(7):073116, 2013.

\bibitem{mourou1981picosecond}
Gerard Mourou, Charles~V Stancampiano, and Daniel Blumenthal.
\newblock Picosecond microwave pulse generation.
\newblock {\em Applied Physics Letters}, 38(6):470--472, 1981.

\bibitem{kim2008drift}
Jungwon Kim, Jonathan~A Cox, Jian Chen, and Franz~X K{\"a}rtner.
\newblock Drift-free femtosecond timing synchronization of remote optical and
  microwave sources.
\newblock {\em Nature Photonics}, 2(12):733--736, 2008.

\bibitem{Ford1998Casimir}
L.~H. Ford.
\newblock Casimir force between a dielectric sphere and a wall: A model for
  amplification of vacuum fluctuations.
\newblock {\em Phys. Rev. A}, 58:4279--4286, Dec 1998.

\bibitem{trenkel2003PatchPotentials}
C.~C. Speake and C.~Trenkel.
\newblock Forces between conducting surfaces due to spatial variations of
  surface potential.
\newblock {\em Phys. Rev. Lett.}, 90:160403, Apr 2003.

\bibitem{nimmrichter2013macroscopicity}
Stefan Nimmrichter and Klaus Hornberger.
\newblock Macroscopicity of mechanical quantum superposition states.
\newblock {\em Physical review letters}, 110(16):160403, 2013.

\bibitem{kovachy2015quantum}
T~Kovachy, P~Asenbaum, C~Overstreet, CA~Donnelly, SM~Dickerson, A~Sugarbaker,
  JM~Hogan, and MA~Kasevich.
\newblock Quantum superposition at the half-metre scale.
\newblock {\em Nature}, 528(7583):530, 2015.

\bibitem{nimmrichter2011testing}
Stefan Nimmrichter, Klaus Hornberger, Philipp Haslinger, and Markus Arndt.
\newblock Testing spontaneous localization theories with matter-wave
  interferometry.
\newblock {\em Physical Review A}, 83(4):043621, 2011.

\bibitem{Biswas:2011ar}
Tirthabir Biswas, Erik Gerwick, Tomi Koivisto, and Anupam Mazumdar.
\newblock {Towards singularity and ghost free theories of gravity}.
\newblock {\em Phys. Rev. Lett.}, 108:031101, 2012.

\bibitem{Biswas:2005qr}
Tirthabir Biswas, Anupam Mazumdar, and Warren Siegel.
\newblock Bouncing universes in string-inspired gravity.
\newblock {\em Journal of Cosmology and Astroparticle Physics}, 2006(03):009,
  2006.

\bibitem{Buoninfante:2017kgj}
Luca Buoninfante, Gaetano Lambiase, and Anupam Mazumdar.
\newblock Quantum solitonic wave-packet of a meso-scopic system in singularity
  free gravity.
\newblock {\em Nuclear Physics B}, 931:250--261, 2018.

\bibitem{Buoninfante:2017rbw}
Luca Buoninfante, Gaetano Lambiase, and Anupam Mazumdar.
\newblock Quantum spreading of a self-gravitating wave-packet in singularity
  free gravity.
\newblock {\em The European Physical Journal C}, 78(1):73, 2018.

\end{thebibliography}
\end{document}